# Surrogate and inverse modeling for two-phase flow in porous media via theory-guided convolutional neural network

Nanzhe Wang[a], Haibin Chang[a,*], and Dongxiao Zhang[b,*]

[a] BIC-ESAT, ERE, and SKLTCS, College of Engineering, Peking University, Beijing 100871, P. R. China
[b] School of Environmental Science and Engineering, Southern University of Science and Technology, Shenzhen 518055, P. R. China

*Corresponding authors: E-mail address: changhaibin@pku.edu.cn (Haibin Chang); zhangdx@sustech.edu.cn (Dongxiao Zhang)

**Abstract**

The theory-guided convolutional neural network (TgCNN) framework, which can incorporate discretized governing equation residuals into the training of convolutional neural networks (CNNs), is extended to two-phase porous media flow problems in this work. The two principal variables of the considered problem, pressure and saturation, are approximated simultaneously with two CNNs, respectively. Pressure and saturation are coupled with each other in the governing equations, and thus the two networks are also mutually conditioned in the training process by the discretized governing equations, which also increases the difficulty of model training. The coupled and discretized equations can provide valuable information in the training process. With the assistance of theory-guidance, the TgCNN surrogates can achieve better accuracy than ordinary CNN surrogates in two-phase flow problems. Moreover, a piecewise training strategy is proposed for the scenario with varying well controls, in which the TgCNN surrogates are constructed for different segments on the time dimension and stacked together to predict solutions for the whole time-span. For scenarios with larger variance of the formation property field, the TgCNN surrogates can also achieve satisfactory performance. The constructed TgCNN surrogates are further used for inversion of permeability fields by combining them with the iterative ensemble smoother (IES) algorithm, and sufficient inversion accuracy is obtained with improved efficiency.

# 1 Introduction

Two-phase flow is very common in subsurface flow problems, such as oil-water flow in oil reservoirs, $CO_2$ injection in geological carbon storage (GCS), etc. Considering the uncertainty involved in subsurface flow problems, inverse modeling is usually necessary while modeling two-phase flows, in which geologic models are calibrated to fit the observation data (Oliver et al., 2008). Subsurface flow models can provide more accurate predictions of future states after being calibrated to the historical data, which would be of great use for oil reservoirs and water resources management. The flow simulation models usually need to be implemented repeatedly in the inverse modeling process. Considering the computational cost demanded, constructing surrogate models that can provide fast approximation of flow simulators is an effective way to improve inverse modeling efficiency. Surrogate modeling methods, including polynomial chaos expansion (PCE) (Chang et al., 2017; Zeng et al., 2011; Zhang et al., 2011), the stochastic collocation method (Li et al., 2009; Liao et al., 2019), Gaussian process (Hamdi et al., 2015; Rana et al., 2018), deep-learning-based methods (Mo et al., 2019; Tang et al., 2020; Zhong et al., 2021), etc., have been widely used in inverse modeling and uncertainty quantification of subsurface flow problems.

Deep-learning-based surrogate models have attracted increasing attention in recent years due to their universal approximation ability and the potential to solve high-dimensional problems. Tripathy and Bilionis (2018) constructed a deep neural network (DNN)-based surrogate for the stochastic elliptic partial differential equation (SPDE), which was further used for uncertainty quantification of diffusion fields. Mo et al. (2019) proposed a deep convolutional encoder-decoder-based surrogate for multiphase flow in a geological carbon storage process, in which the binary cross entropy (BCE) loss of the binarized saturation field is incorporated to better approximate the discontinuous saturation front. Zhong et al. (2021) constructed a surrogate model based on the coupled generative adversarial network (Co-GAN), which can forecast reservoir pressure and fluid saturation simultaneously under reservoir property uncertainty with a single model. Despite the success achieved with these deep-learning-based surrogates, most of them are purely data-driven, which demands a large amount of training data, and the physical principles of the studied problems are usually ignored. Therefore, an increasing number of researchers are attempting to incorporate physical theories while constructing deep-learning-based surrogate models.

The physics-informed neural network (PINN) proposed by Raissi et al. (2019) is a well-known framework, which utilizes the residuals of governing equations to regulate the deep neural network, and the requisite partial differentials can be obtained via automatic differentiation (AD). Wang et al. (2020) proposed a theory-guided neural network, in which, besides the governing equations, expert knowledge and engineering controls are also incorporated into the training process. Karumuri et al. (2020) developed a simulator-free surrogate based on the deep residual network (ResNet), in which the physics-informed loss was derived from variational principles. The trained surrogate is further used for solving uncertainty propagation and inverse problems of elliptic stochastic partial differential equations (SPDEs). Wang et al. (2021c) proposed a theory-guided neural network surrogate for subsurface flow problems, and constructed a composite model in their work, which can be employed for cases with varying boundary conditions and field variances. Zhu et al. (2019) developed a physics-constrained deep learning surrogate model, which uses a convolutional encoder-decoder architecture, and the Sobel filter is adopted to approximate the spatial gradients. These theory-guided or physics-constrained deep learning models can also be applied in reservoir simulation. In our former work, a theory-guided convolutional neural network (TgCNN) is constructed for uncertainty quantification and data assimilation of single-phase flow problems in oil reservoirs (Wang et al., 2021b), which incorporated discretized governing equations with finite difference (FD) into the loss function to impose the physical constraints. However, only single-phase flow problems in reservoir simulation are investigated in this work for proof-of-concept.

In this work, the TgCNN framework is extended to two-phase flow problems, which constitutes an important step towards practical application. Oil-water flow in water-flooding problems is considered, in which pressure and saturation are the principle variables. Two convolutional neural networks (CNNs) are constructed to approximate pressure and saturation, respectively. The two networks are also coupled with each other in the training process by the governing equation residuals, which also increases the difficulty of model training. With the theory guidance, the TgCNN surrogates can achieve better accuracy than ordinary CNN surrogates. In addition, a scenario with varying well controls is also investigated, in which the TgCNN surrogates are constructed in a piecewise manner on the time dimension and stacked together to predict solutions for the whole time-span. The performance of TgCNN for scenarios with larger variance of formation property fields is also investigated. Moreover, the transfer learning strategy is adopted to accelerate the training process for the new-variance case. The trained TgCNN surrogates are further used for inverse modeling by combining them with the iterative ensemble smoother (IES) algorithm, and improved inversion efficiency is achieved with sufficient estimation accuracy.

The remainder of this paper is organized as follows. In section 2, the governing equations of subsurface two-phase problems are introduced, and the finite difference (FD) method used to discretize the equations is also illustrated. In section 3, the framework of TgCNN for two-phase flow problems is first introduced, and then the TgCNN surrogate-based iterative ensemble smoother (IES) is introduced for inverse modeling. In section 4, a two-dimensional water-flooding case is considered to test the performance of the TgCNN framework for two-phase problems. In section 5, the constructed TgCNN surrogates are utilized for inverse modeling of the designed case. Finally, discussions and conclusions are given in section 6.

## 2 Governing Equations

In this work, two-phase flow problems in oil reservoirs are considered. The governing equation of oil (or water) phase in two-dimensional reservoirs is presented as follows:

$$\frac{\partial}{\partial x}\left(\rho_l \frac{k_x k_{rl}}{\mu_l} \frac{\partial P_l}{\partial x}\right) + \frac{\partial}{\partial y}\left(\rho_l \frac{k_y k_{rl}}{\mu_l} \frac{\partial P_l}{\partial y}\right) + q_l = \frac{\partial}{\partial t}\left(\phi \rho_l S_l\right), \quad l = w, o \quad (1)$$

where $l$ denotes the phase of the liquid, which can be $o$ for the oil phase and $w$ for the water phase; $\rho_l$ and $\mu_l$ denote the density and viscosity of the $l$ phase, respectively; $P_l$ and $S_l$ denote the pressure and saturation of the $l$ phase, respectively, which are the quantities of interest in this problem; $k_x$ and $k_y$ denote the absolute permeability in x- and y-direction, respectively; $k_{rl}$ denotes the relative permeability of the $l$ phase, and it is worth noting that the relative permeability $k_{rl}$ is a nonlinear function of saturation $S_l$, i.e., $k_{rl}(S_l)$, which increases the nonlinearity of the governing equation; $\phi$ denotes the porosity of the reservoir; and $q_l$ denotes the sink/source terms in the reservoir, which are usually wells and can be represented with Peacemen's well model in numerical simulation (Peaceman, 1983):

$$q_{li,j} = \frac{2\pi \left(k_x k_y\right)_{i,j}^{1/2} \rho_{sc} \Delta z}{\ln(r_o / r_w)} \left(\frac{k_{rl}}{\mu_l}\right)(P_{li,j} - P_{wf}) \tag{2}$$

where $\Delta z$ denotes the thickness of the grid block; $\rho_{sc}$ denotes the density of oil at standard conditions; $P_{li,j}$ denotes the pressure of phase $l$ at well block $(i,j)$; $P_{wf}$ denotes the bottom hole pressure (BHP); $r_w$ denotes the radius of the wellbore; and $r_o$ denotes the equivalent radius of the well block, which can be calculated with (Peaceman, 1983):

$$r_o = 0.28 \frac{\left[\left(k_y / k_x\right)^{1/2} \Delta x^2 + \left(k_x / k_y\right)^{1/2} \Delta y^2\right]^{1/2}}{\left(k_y / k_x\right)^{1/4} + \left(k_x / k_y\right)^{1/4}} \tag{3}$$

Some auxiliary equations are also needed to constrain the quantities:

$$S_o + S_w = 1 \tag{4}$$

$$P_{cow} = P_o - P_w = f(S_w) \tag{5}$$

where $P_{cow}$ denotes the capillary pressure between the oil and water phase; and $f(S_w)$ denotes the relationship between the water saturation and capillary pressure.

If the compressibility of the rock is ignored, the right term of Eq. (1) can be decomposed into:

$$\frac{\partial}{\partial t}(\phi \rho_l S_l) = \phi S_l \frac{\partial \rho_l}{\partial t} + \phi \rho_l \frac{\partial S_l}{\partial t} \tag{6}$$

And Eq. (6) can further become:

$$\frac{\partial}{\partial t}(\phi \rho_l S_l) = \phi C_l S_l \rho_l \frac{\partial P_l}{\partial t} + \phi \rho_l \frac{\partial S_l}{\partial t} \tag{7}$$

by incorporating $C_l = -\frac{1}{V}\frac{\partial V}{\partial P_l} = \frac{1}{\rho_l}\frac{\partial \rho_l}{\partial P_l}$, where $C_l$ denotes the compressibility of phase $l$.

Therefore, Eq. (1) can be rewritten as:

$$\frac{\partial}{\partial x}\left(\rho_l \frac{k_x k_{rl}}{\mu_l}\frac{\partial P_l}{\partial x}\right) + \frac{\partial}{\partial y}\left(\rho_l \frac{k_y k_{rl}}{\mu_l}\frac{\partial P_l}{\partial y}\right) + q_l = \phi C_l S_l \rho_l \frac{\partial P_l}{\partial t} + \phi \rho_l \frac{\partial S_l}{\partial t}. \tag{8}$$

To solve the problem numerically, the finite difference (FD) method is usually utilized, in which the governing equation can be discretized as follows:

$$\frac{\left(\rho_l \frac{k_x k_{rl}}{\mu_l}\right)^{n+1}_{i+1/2,j}\frac{P^{n+1}_{li+1,j}-P^{n+1}_{li,j}}{(1/2)(\Delta x_i + \Delta x_{i+1})} - \left(\rho_l \frac{k_x k_{rl}}{\mu_l}\right)^{n+1}_{i-1/2,j}\frac{P^{n+1}_{li,j}-P^{n+1}_{li-1,j}}{(1/2)(\Delta x_i + \Delta x_{i-1})}}{\Delta x_i} +$$

$$\frac{\left(\rho_l \frac{k_y k_{rl}}{\mu_l}\right)^{n+1}_{i,j+1/2}\frac{P^{n+1}_{li,j+1}-P^{n+1}_{li,j}}{(1/2)(\Delta y_j + \Delta y_{j+1})} - \left(\rho_l \frac{k_y k_{rl}}{\mu_l}\right)^{n+1}_{i,j-1/2}\frac{P^{n+1}_{li,j}-P^{n+1}_{li,j-1}}{(1/2)(\Delta y_j + \Delta y_{j-1})}}{\Delta y_j} + \tag{9}$$

$$q_{li,j} = \phi_{i,j} C_l S^n_{li,j} \rho_l \frac{P^{n+1}_{li,j}-P^n_{li,j}}{\Delta t} + \phi_{i,j} \rho_l \frac{S^{n+1}_{li,j}-S^n_{li,j}}{\Delta t}$$

where $i$, $j$, and $n$ denote the index of x-coordinates, y-coordinates, and time-coordinates, respectively; and $\left(\rho_l \frac{k_x k_{rl}}{\mu_l}\right)^{n+1}_{i+1/2,j}$ can be calculated with the harmonic mean of two adjacent grids and in an upstream weighted manner, as presented below:

$$\left(\rho_l \frac{k_x k_{rl}}{\mu_l}\right)^{n+1}_{i+1/2,j} = k_{x,i+1/2,j}\left[\frac{\rho_l(P^n_l)\cdot k_{rl}(S^n_l)}{\mu_l(P^n_l)}\right]_{m,j} \tag{10}$$

where $k_{x,i+1/2,j} = \frac{\Delta x_{i+1} + \Delta x_i}{\frac{\Delta x_{i+1}}{k_{x,i+1,j}} + \frac{\Delta x_i}{k_{x,i,j}}}$, and $m = \begin{cases} i, & P_{l,i,j} > P_{l,i+1,j} \\ i+1, & P_{l,i,j} < P_{l,i+1,j} \end{cases}$.

Then, the discretized nonlinear partial differential equations can be used to formulate the linear equation groups and solved with numerical methods, such as the implicit pressure explicit saturation procedure (IMPES), fully implicit procedure, etc.

## 3 Methodology

In this section, the theory-guided convolutional neural network (TgCNN) and iterative ensemble smoother (IES) for data assimilation in two-phase flow problems are presented. The TgCNN can be used for surrogate modeling to improve the efficiency of forward calculation, and the IES can used for parameter inversion.

### 3.1 Theory-guided convolutional neural network (TgCNN) for two-phase flow

The TgCNN framework was proposed in our previous work (Wang et al., 2021b), while in this work, the TgCNN framework is extended to two-phase flow problems. Considering that there are two principal variables in this problem, i.e., pressure and saturation, two convolutional encoder-decoders are constructed to approximate the quantities of interest, respectively, i.e.:

$$\hat{P} = \mathrm{N}_P(K,T;\theta_P) \tag{11}$$

$$\hat{S} = \mathrm{N}_S(K,T;\theta_S) \tag{12}$$

where $K$ denotes the image of permeability fields, and $T$ denotes the time matrix, as shown in **Figure 1**; $\mathrm{N}_P$ and $\mathrm{N}_S$ denote the mapping relationships approximated with two convolutional encoder-decoders from the model parameters ($K,T$) to the quantities of interest; and $\theta_P$ and $\theta_S$ denote the parameters (weights and bias) of the convolutional networks.

With the training data from reservoir simulators, the two convolutional networks can be trained in the conventional data-driven way:

$$\begin{aligned} L_{\mathrm{data}}(\theta_P, \theta_S) &= \frac{1}{N_r N_t}\sum_{i=1}^{N_r}\sum_{j=1}^{N_t}\left\|\hat{P}^{K_i,T_j} - P^{K_i,T_j}\right\|_2^2 + \frac{1}{N_r N_t}\sum_{i=1}^{N_r}\sum_{j=1}^{N_t}\left\|\hat{S}^{K_i,T_j} - S^{K_i,T_j}\right\|_2^2 \\ &= \frac{1}{N_r N_t}\sum_{i=1}^{N_r}\sum_{j=1}^{N_t}\left\|\mathrm{N}_P(K_i,T_j;\theta_P) - P^{K_i,T_j}\right\|_2^2 + \frac{1}{N_r N_t}\sum_{i=1}^{N_r}\sum_{j=1}^{N_t}\left\|\mathrm{N}_S(K_i,T_j;\theta_S) - S^{K_i,T_j}\right\|_2^2 \end{aligned}, \tag{13}$$

where $N_r$ denotes the number of permeability field realizations solved with reservoir simulators and used as the labeled training datasets; and $N_t$ denotes the number of time-steps in each realization.

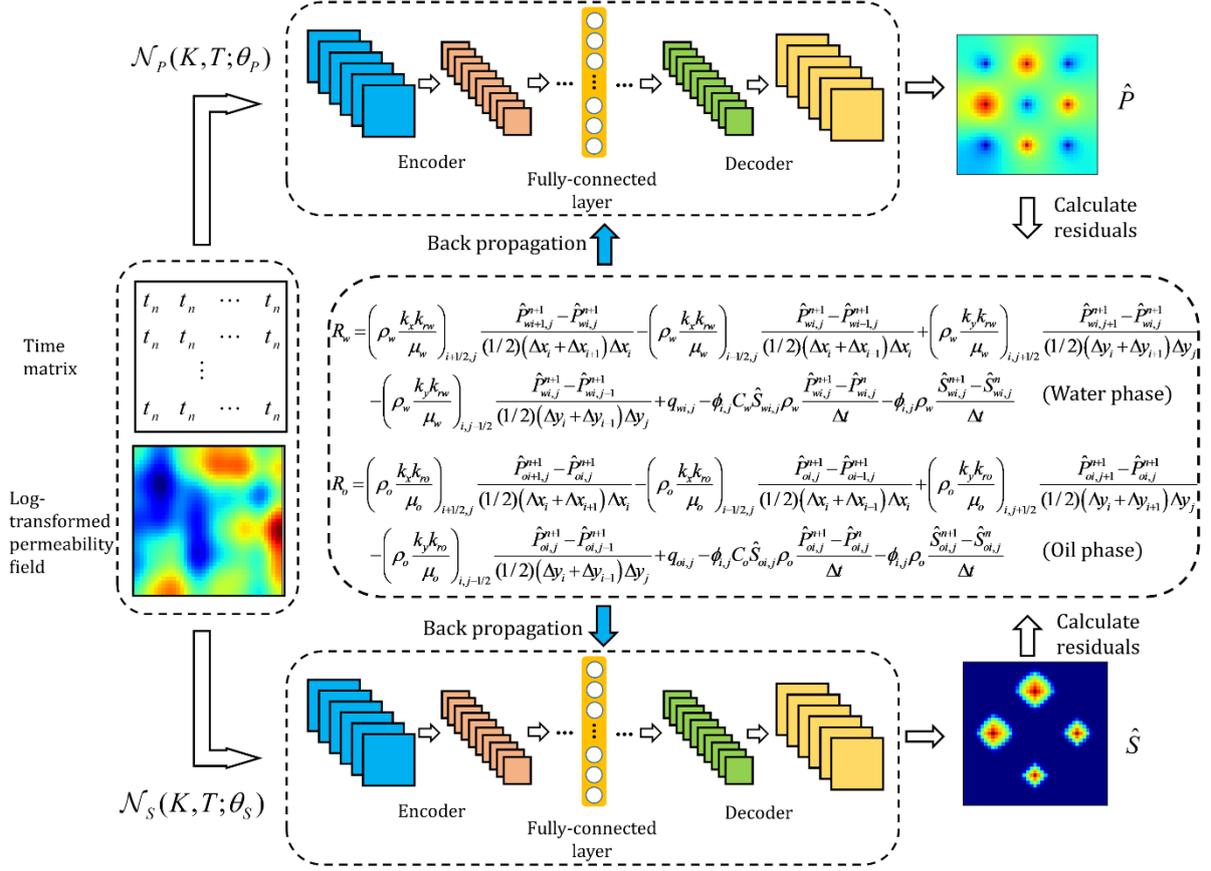

**Figure 1.** Framework of TgCNN for two-phase flow.

In the TgCNN framework, the governing equations of the studied problems can be incorporated into the training process of the networks to achieve theory-guided training and improve model prediction accuracy (Wang et al., 2021b). Here, for the two-phase flow problem, the governing equations can also be transformed into the residual form as follows:

$$R_w(K,T;\ \theta_P,\theta_S) = \frac{\left(\rho_w \frac{k_x k_{rw}}{\mu_w}\right)_{i+1/2,j} \frac{\hat{P}_{wi+1,j}^{n+1} - \hat{P}_{wi,j}^{n+1}}{(1/2)(\Delta x_i + \Delta x_{i+1})} - \left(\rho_w \frac{k_x k_{rw}}{\mu_w}\right)_{i-1/2,j} \frac{\hat{P}_{wi,j}^{n+1} - \hat{P}_{wi-1,j}^{n+1}}{(1/2)(\Delta x_i + \Delta x_{i-1})}}{\Delta x_i} +$$
$$\frac{\left(\rho_w \frac{k_y k_{rw}}{\mu_w}\right)_{i,j+1/2} \frac{\hat{P}_{wi,j+1}^{n+1} - \hat{P}_{wi,j}^{n+1}}{(1/2)(\Delta y_j + \Delta y_{j+1})} - \left(\rho_w \frac{k_y k_{rw}}{\mu_w}\right)_{i,j-1/2} \frac{\hat{P}_{wi,j}^{n+1} - \hat{P}_{wi,j-1}^{n+1}}{(1/2)(\Delta y_j + \Delta y_{j-1})}}{\Delta y_j} + \quad (14)$$
$$q_{wi,j} - \phi_{i,j} C_w \hat{S}_{wi,j} \rho_w \frac{\hat{P}_{wi,j}^{n+1} - \hat{P}_{wi,j}^n}{\Delta t} - \phi_{i,j} \rho_w \frac{\hat{S}_{wi,j}^{n+1} - \hat{S}_{wi,j}^n}{\Delta t}$$

$$R_o(K,T;\theta_P,\theta_S) = \frac{\left(\rho_o\frac{k_x k_{ro}}{\mu_o}\right)_{i+1/2,j}\frac{\hat{P}_{oi+1,j}^{n+1}-\hat{P}_{oi,j}^{n+1}}{(1/2)(\Delta x_i+\Delta x_{i+1})}-\left(\rho_o\frac{k_x k_{ro}}{\mu_o}\right)_{i-1/2,j}\frac{\hat{P}_{oi,j}^{n+1}-\hat{P}_{oi-1,j}^{n+1}}{(1/2)(\Delta x_i+\Delta x_{i-1})}}{\Delta x_i}+$$

$$\frac{\left(\rho_o\frac{k_y k_{ro}}{\mu_o}\right)_{i,j+1/2}\frac{\hat{P}_{oi,j+1}^{n+1}-\hat{P}_{oi,j}^{n+1}}{(1/2)(\Delta y_i+\Delta y_{i+1})}-\left(\rho_o\frac{k_y k_{ro}}{\mu_o}\right)_{i,j-1/2}\frac{\hat{P}_{oi,j}^{n+1}-\hat{P}_{oi,j-1}^{n+1}}{(1/2)(\Delta y_i+\Delta y_{i-1})}}{\Delta y_j}+ \quad (15)$$

$$q_{oi,j}-\phi_{i,j}C_o\hat{S}_{oi,j}\rho_o\frac{\hat{P}_{oi,j}^{n+1}-\hat{P}_{oi,j}^{n}}{\Delta t}-\phi_{i,j}\rho_o\frac{\hat{S}_{oi,j}^{n+1}-\hat{S}_{oi,j}^{n}}{\Delta t}$$

where $\hat{P}_o = \hat{P}_w + P_{cow}$, and $\hat{S}_o = 1 - \hat{S}_w$. It can be seen that the pressure and saturation are coupled together for the two-phase flow problem, which increases the difficulty of model training. The residuals of the governing equations can then constitute the physics-constrained loss function and be minimized:

$$L_{\text{GE}}(\theta_P,\theta_S) = \frac{1}{N_{vr}N_t}\sum_{i=1}^{N_{vr}}\sum_{j=1}^{N_t}\|R_w(K_i,T_j;\theta_P,\theta_S)\|_2^2 + \frac{1}{N_{vr}N_t}\sum_{i=1}^{N_{vr}}\sum_{j=1}^{N_t}\|R_o(K_i,T_j;\theta_P,\theta_S)\|_2^2 \quad (16)$$

where $N_{vr}$ denotes the total number of realizations used to impose the physical constraints, which can be termed virtual realizations (Wang et al., 2021b, d). Furthermore, the boundary conditions and initial conditions can also be discretized and transformed into the residual forms in a similar manner, and thus the total loss function of the TgCNN model for the two-phase flow problem can be formulated as:

$$L(\theta_P,\theta_S) = \lambda_{\text{data}}L_{\text{data}}(\theta_P,\theta_S) + \lambda_{\text{GE}}L_{\text{GE}}(\theta_P,\theta_S) + \lambda_{\text{BC}}L_{\text{BC}}(\theta_P,\theta_S) + \lambda_{\text{IC}}L_{\text{IC}}(\theta_P,\theta_S) \quad (17)$$

where $L_{\text{BC}}$ and $L_{\text{IC}}$ denote the loss functions of the boundary and initial conditions, respectively; and the hyperparameter $\lambda$ denotes the weight of each term in the total loss function. For the initial conditions and Dirichlet boundary conditions, the constraints can be imposed as 'hard constraints' or 'soft constraints', which are much easier to implement (Wang et al., 2021b, c). For the Neumann boundary conditions, however, the equations can be discretized with the FD method and imposed in a manner similar to the governing equations. Additional details can be found in our previous work (Wang et al., 2021d).

By minimizing the loss function with various optimization algorithms, such as stochastic gradient descent (SGD) (Bottou, 2010) and adaptive moment estimation (Adam) (Kingma &

Ba, 2015), the TgCNN models can be trained and further used for predicting solutions of new permeability realizations. Considering that forward calculation with the convolutional network is much faster than running the numerical reservoir simulators, the data assimilation process can be sped up with the trained TgCNN surrogates.

**3.2 TgCNN-based iterative ensemble smoother (IES)**

The trained TgCNN surrogates can then be combined with the iterative ensemble smoother (IES) for inverse modeling. Inverse modeling is a crucial part for modeling flow in porous media, and IES is an effective ensemble-based method for implementing it. Estimating the formation property field by assimilating the historical data to reduce its uncertainty and to make more accurate reservoir production predictions are the main purposes of reservoir inverse modeling. This can be viewed as an optimization problem, as the objective function shown below (Oliver et al., 2008):

$$O(\mathbf{m}) = \frac{1}{2}\left(\mathbf{d}^{obs} - g(\mathbf{m})\right)^T C_D^{-1}\left(\mathbf{d}^{obs} - g(\mathbf{m})\right) \\ + \frac{1}{2}\left(\mathbf{m} - \mathbf{m}^{pr}\right)^T C_M^{-1}\left(\mathbf{m} - \mathbf{m}^{pr}\right) \tag{18}$$

where $\mathbf{m}$ denotes the model parameters to be estimated, which refers to the permeability field in this problem; $\mathbf{m}^{pr}$ denotes the prior estimation of model parameters; $\mathbf{d}^{obs}$ denotes the production measurements; $C_D$ and $C_M$ denote the covariance matrix of measured data and model parameters, respectively; and $g(\mathbf{m})$ denotes the forward calculation of reservoir simulator with model parameters $\mathbf{m}$.

The optimization problem Eq. (18) can be solved with the Levenberg–Marquardt (LM) algorithm, in which the model parameters $\mathbf{m}$ are updated iteratively (Chen & Oliver, 2013):

$$\begin{aligned}\mathbf{m}_{l+1} &= \mathbf{m}_l - \left[(1+\lambda_l)C_M^{-1} + G_l^T C_D^{-1} G_l\right]^{-1}\left[C_M^{-1}\left(\mathbf{m}_l - \mathbf{m}^{pr}\right) + G_l^T C_D^{-1}\left(g(\mathbf{m}_l) - \mathbf{d}^{obs}\right)\right] \\ &= \mathbf{m}_l - \frac{1}{1+\lambda_l}\left[C_M - C_M G_l^T\left((1+\lambda_l)C_D + G_l C_M G_l^T\right)^{-1} G_l C_M\right] C_M^{-1}\left(\mathbf{m}_l - \mathbf{m}^{pr}\right) \\ &\quad - C_M G_l^T\left((1+\lambda_l)C_D + G_l C_M G_l^T\right)^{-1}\left(g(\mathbf{m}_l) - \mathbf{d}^{obs}\right)\end{aligned} \tag{19}$$

where $l$ denotes the index of iteration number; $\lambda_l$ denotes the multiplier, which can adjust the effect of data mismatch in the objective function; and $G_l$ denotes the sensitivity matrix of production data with respect to the model parameters, which is usually challenging to calculate directly for subsurface flow problems. Therefore, the ensemble-based methods can be adopted to avoid direct calculation of sensitivity matrix $G_l$. In the ensemble-based method, each realization in the ensemble should be updated, and thus Eq. (19) can be revised as follows:

$$\begin{aligned}\mathbf{m}_{l+1,j} = \mathbf{m}_{l,j} - \frac{1}{1+\lambda_l}\left[C_{M_l} - C_{M_l}\bar{G}_l^T\left((1+\lambda_l)C_D + \bar{G}_l C_{M_l}\bar{G}_l^T\right)^{-1}\bar{G}_l C_{M_l}\right]C_M^{-1}\left(\mathbf{m}_{l,j} - \mathbf{m}_j^{pr}\right) \\ - C_{M_l}\bar{G}_l^T\left((1+\lambda_l)C_D + \bar{G}_l C_{M_l}\bar{G}_l^T\right)^{-1}\left(g(\mathbf{m}_{l,j}) - \mathbf{d}_j^{obs}\right), \quad j=1,\ldots,N_e\end{aligned} \quad (20)$$

where $j$ denotes the index of realizations; $\bar{G}_l$ denotes the average sensitivity matrix of different realizations; and $N_e$ denotes the total number of realizations in the ensemble. In addition, the following approximations can be adopted to use the information from the ensemble and simplify the calculation (Chang et al., 2017; Li & Reynolds, 2009):

$$C_{M_l}\bar{G}_l^T \approx C_{M_l D_l} \quad (21)$$

$$\bar{G}_l C_{M_l}\bar{G}_l^T \approx C_{D_l D_l}, \quad (22)$$

where $C_{M_l D_l}$ denotes the cross covariance matrix between the model parameters and measurements; and $C_{D_l D_l}$ denotes the covariance matrix of measurements. Then, Eq. (20) can be rewritten as:

$$\begin{aligned}\mathbf{m}_{l+1,j} = \mathbf{m}_{l,j} - \frac{1}{1+\lambda_l}\left[C_{M_l} - C_{M_l D_l}\left((1+\lambda_l)C_D + C_{D_l D_l}\right)^{-1} C_{D_l M_l}\right]C_M^{-1}\left(\mathbf{m}_{l,j} - \mathbf{m}_j^{pr}\right) \\ - C_{M_l D_l}\left((1+\lambda_l)C_D + C_{D_l D_l}\right)^{-1}\left(g(\mathbf{m}_{l,j}) - \mathbf{d}_j^{obs}\right), \quad j=1,\ldots,N_e\end{aligned}. \quad (23)$$

It can be seen that in Eq. (23), the reservoir simulators need to be run iteratively during the data assimilation process, which would be both computationally expensive and time-consuming. The efficiency of the data assimilation process can be improved by using the trained TgCNN surrogate to replace the simulators. The surrogate-based IES has also been used in Chang et al. (2017) and Wang et al. (2021a) for inverse modeling. The updating scheme of

model parameters with the TgCNN surrogate can then be formulated as:

$$\mathbf{m}_{l+1,j} = \mathbf{m}_{l,j} - \frac{1}{1+\lambda_l}\left[C_{M_l} - C_{M_lD_l}^{surr}\left((1+\lambda_l)C_D + C_{D_lD_l}^{surr}\right)^{-1}C_{D_lM_l}^{surr}\right]C_M^{-1}\left(\mathbf{m}_{l,j} - \mathbf{m}_j^{pr}\right)$$
$$- C_{M_lD_l}^{surr}\left((1+\lambda_l)C_D + C_{D_lD_l}^{surr}\right)^{-1}\left(g^{surr}(\mathbf{m}_{l,j}) - \mathbf{d}_j^{obs}\right), \quad j=1,\ldots,N_e \quad (24)$$

where superscript '*surr*' denotes the results calculated from surrogate models; and $C_{M_lD_l}^{surr}$ and $C_{D_lD_l}^{surr}$ can be calculated with:

$$C_{M_lD_l}^{surr} = \frac{1}{N_e-1}\sum_{j=1}^{N_e}\left\{\left[\mathbf{m}_{l,j} - <\mathbf{m}_{l,j}>\right] \times \left[g^{surr}(\mathbf{m}_{l,j}) - <g^{surr}(\mathbf{m}_{l,j})>\right]^T\right\} \quad (25)$$

$$C_{D_lD_l}^{surr} = \frac{1}{N_e-1}\sum_{j=1}^{N_e}\left\{\left[g^{surr}(\mathbf{m}_{l,j}) - <g^{surr}(\mathbf{m}_{l,j})>\right] \times \left[g^{surr}(\mathbf{m}_{l,j}) - <g^{surr}(\mathbf{m}_{l,j})>\right]^T\right\}. \quad (26)$$

Therefore, with IES, the direct calculation of the sensitivity matrix can be avoided, and the efficiency of the iteration process can also be improved by using the TgCNN surrogate for forward calculation.

## 4 Surrogate Modeling with TgCNN

In this section, the TgCNN framework is utilized for surrogate modeling of water flooding two-phase flow problems in oil reservoirs.

### 4.1 Model descriptions

Consider a two-dimensional reservoir, and the domain is a square with side length being 820 m, which is discretized into $41 \times 41$ grid blocks. The four boundaries of the reservoir are closed boundaries (no-flow). The reservoir is located at a depth of 3000 m, and the thickness of the reservoir is set to be 20 m. The initial and reference pressure at reference depth 3010 m is set to be 300 bar, and the porosity is assumed to be constant, which takes a value of 0.15. There are five producing wells for oil production and four injection wells for water flooding in the reservoir, as shown in **Figure 2**. Both the producing wells and injection wells are operated to follow bottom hole pressure (BHP) constraints. The producing wells product with BHP being 280 bar and the injection wells inject water with BHP being 320 bar. The density, viscosity, compressibility, and formation volume factor of oil at reference pressure are 1000 kg/m$^3$, 0.32 cP, $6\times10^{-5}$ 1/bar, and 1.0 $rm^3/sm^3$, respectively. Those properties of the oil phase are 800 kg/m$^3$, 0.85 cP, $9\times10^{-5}$ 1/bar, and 1.1 $rm^3/sm^3$, respectively. To simplify the problem, capillary pressure is neglected, and thus $P_w = P_o$ in this work. The relative permeability of the oil and water phase is a nonlinear function of water saturation, which can be calculated as follows:

$$k_{ro}(S_w) = k_{ro}^0 \left( \frac{1 - S_w - S_{or}}{1 - S_{wr} - S_{or}} \right)^a, \tag{27}$$

$$k_{rw}(S_w) = k_{rw}^0 \left( \frac{S_w - S_{wr}}{1 - S_{wr} - S_{or}} \right)^b \tag{28}$$

where $k_{ro}^0 = 1.0$, $k_{rw}^0 = 0.8$, $S_{or} = 0.2$, $S_{wr} = 0.2$, $a = 3$, and $b = 2$ in this case. The relative permeability curves of the oil and water phase are presented in **Figure 3**.

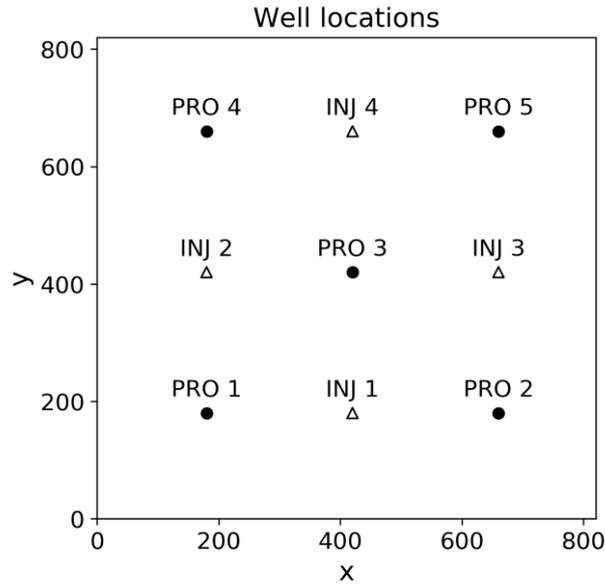

**Figure 2.** Well locations of the water-flooding problem. The filled circles denote the producers and the empty triangles denote the injectors.

The permeability field of the reservoir is assumed to be a log-normal stochastic field, which conforms to the following statistics:

$$\langle \ln K \rangle = 4.0 \quad \ln(\text{mD}), \tag{29}$$

$$\sigma_{\ln K} = 0.5, \tag{30}$$

$$C_{\ln K}(\mathbf{x}_1, \mathbf{x}_2) = \sigma_{\ln K}^2 \exp\left[-\left(\frac{|x_1 - x_2|}{\eta_x} + \frac{|y_1 - y_2|}{\eta_y}\right)\right], \tag{31}$$

where $\langle \ln K \rangle$ and $\sigma_{\ln K}$ denote the mean and standard deviation of the stochastic field, respectively; $C_{\ln K}$ denotes the covariance function of stochastic field $\ln K$; and $\eta_{x \text{ (or } y)}$ denotes the correlation length in the x (or y) direction of the stochastic field, which takes the value of $\eta_x = \eta_y = 328 \text{ m}$ in this case. Then, the Karhunen–Loeve expansion (KLE) can be adopted to generate the permeability field realizations following those statistics (Li & Zhang, 2007). In this work, 85% of the information of the stochastic field is maintained to generate the permeability fields, resulting in 32 truncated terms in KLE. The two-phase flow problem can be solved with the reservoir simulator Eclipse for providing training data, which can also be used as a comparison reference for the TgCNN surrogates (Schlumberger, 2009).

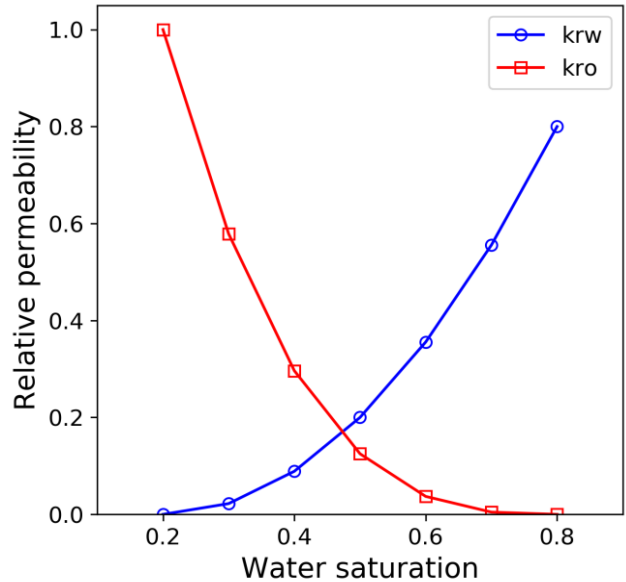

**Figure 3.** Relative permeability curves of the oil and water phase.

## 4.2 TgCNN versus CNN

In this subsection, TgCNN surrogates are constructed for the two-phase flow problem. To construct the training dataset, 80 permeability field realizations are generated with KLE and solved with the simulator Eclipse. To impose the physical constraints in the training process,

200 virtual realizations are generated, which are used to calculate the $L_{\text{GE}}(\theta_P, \theta_S)$ term in Eq. (17). In this case, the total simulation time of the problem is 100 d, which is divided into 100 time-steps with $\Delta t = 1 \text{ d}$. Two CNNs are constructed for approximating the distribution of pressure and water saturation, respectively. The two networks are trained simultaneously, and the total training process takes approximately 5.50 h (19807.55 s) on the NVIDIA TITAN RTX GPU card. To test the accuracy of the trained TgCNN surrogates, 200 testing realizations are generated and solved with the simulator for comparison.

The comparison between predictions from TgCNN surrogates and the reference from the simulator of three sampled realizations are presented in **Figure 4**. It can be seen that the predictions of TgCNN are relatively accurate for both pressure and water saturation distribution. In addition, the scatter plots of predictions and reference values for the 200 testing realizations are presented in **Figure 5** (a) and (b), which also demonstrate the accuracy of the trained TgCNN surrogates. In order to investigate the effectiveness of the theory-guidance (physical constraints) incorporated during the training process, two purely data-driven CNNs are also trained for the pressure and saturation, respectively. The scatter plots of reference values and predictions from CNN models for pressure and saturation are presented in **Figure 5** (c) and (d). It is obvious that the performance of CNN surrogates is inferior to that of TgCNN surrogates, which further shows the effectiveness of the physical constraints (Wang et al., 2021d).

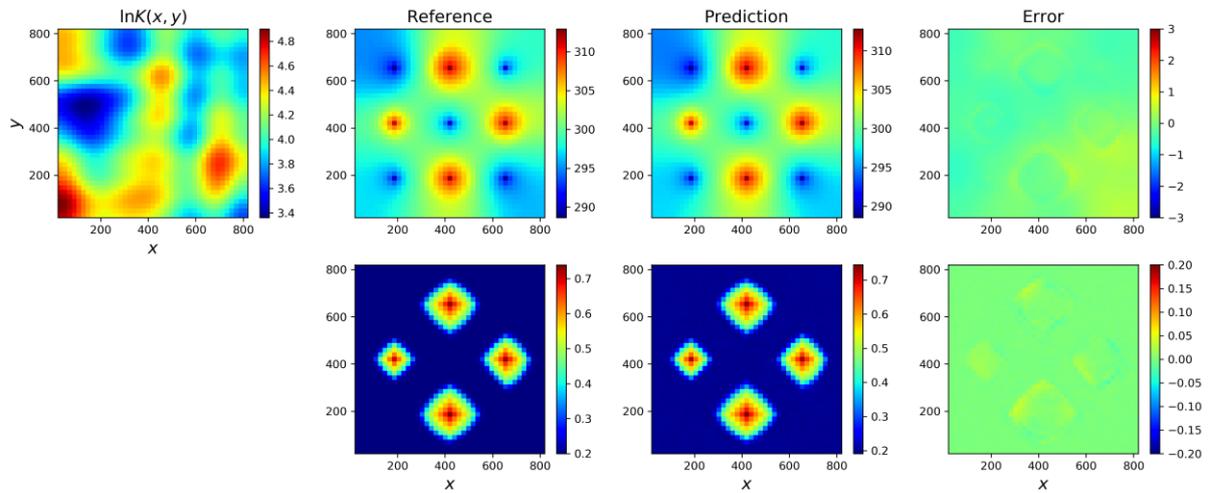

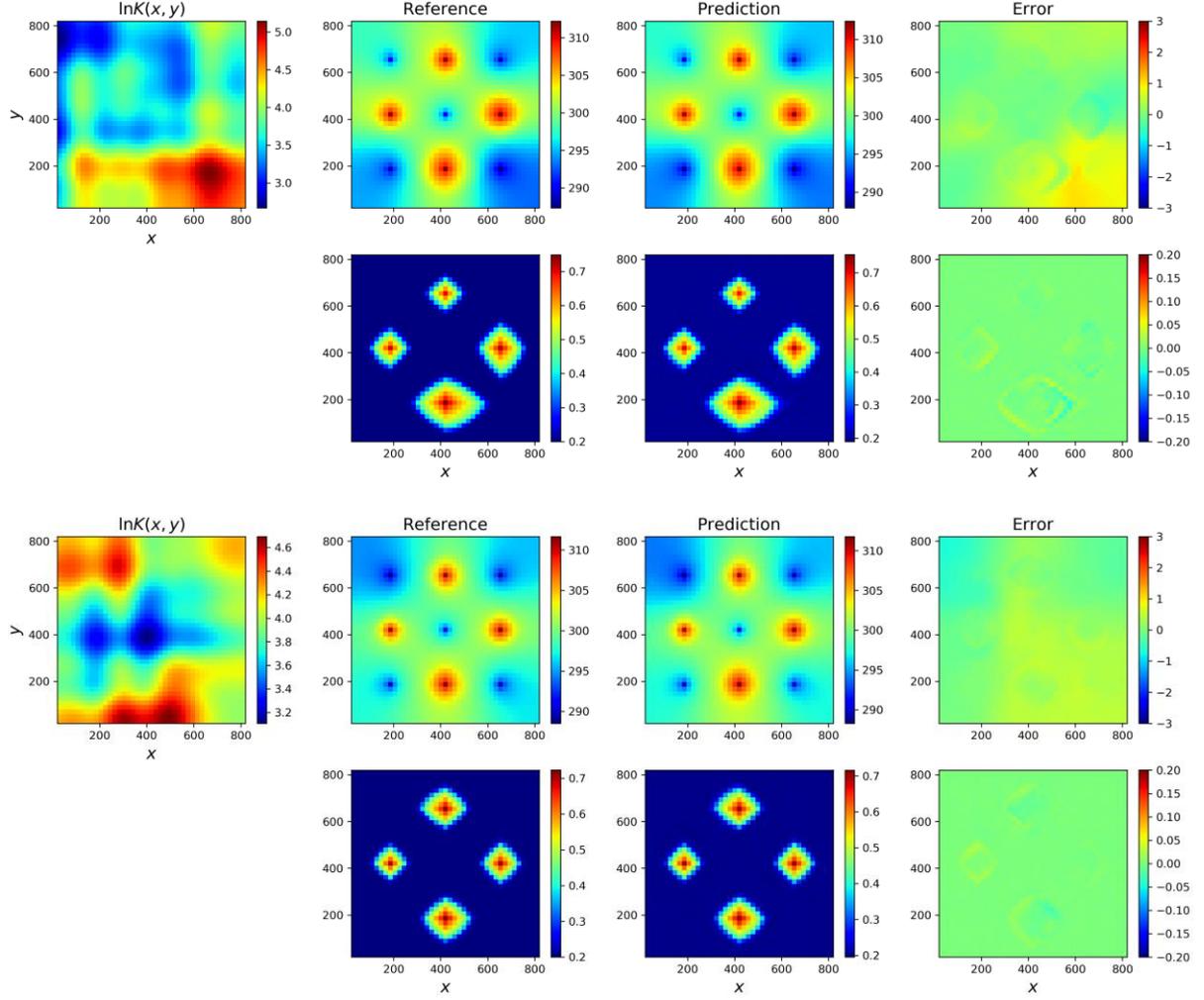

**Figure 4.** Predictions and references of pressure and saturation distribution for three different permeability fields.

In order to compare the performance of CNN and TgCNN quantitatively, the relative $L_2$ error and $R^2$ score are utilized, which can be calculated with:

$$L_2 = \frac{\left\| P_{\text{pred}} - P_{\text{ref}} \right\|_2}{\left\| P_{\text{ref}} \right\|_2} \tag{32}$$

$$R^2 = 1 - \frac{\sum_{n=1}^{N_{\text{gird}}} (P_{\text{pred},n} - P_{\text{ref},n})^2}{\sum_{n=1}^{N_{\text{grid}}} (P_{\text{ref},n} - \overline{P}_{\text{ref}})^2} \tag{33}$$

where $P_{\text{pred}}$ and $P_{\text{ref}}$ denote the predictions and reference values, respectively; $N_{\text{grid}}$ denotes the total number of grid blocks in the domain; and $\overline{P}_{\text{ref}}$ denotes the mean of $P_{\text{ref}}$.

The histograms of relative $L_2$ error and $R^2$ score of the predictions from TgCNN surrogates and CNN surrogates are presented in **Figure 6**. It can be seen that the TgCNN surrogates obtain higher accuracy than the CNN surrogates statistically and quantitatively.

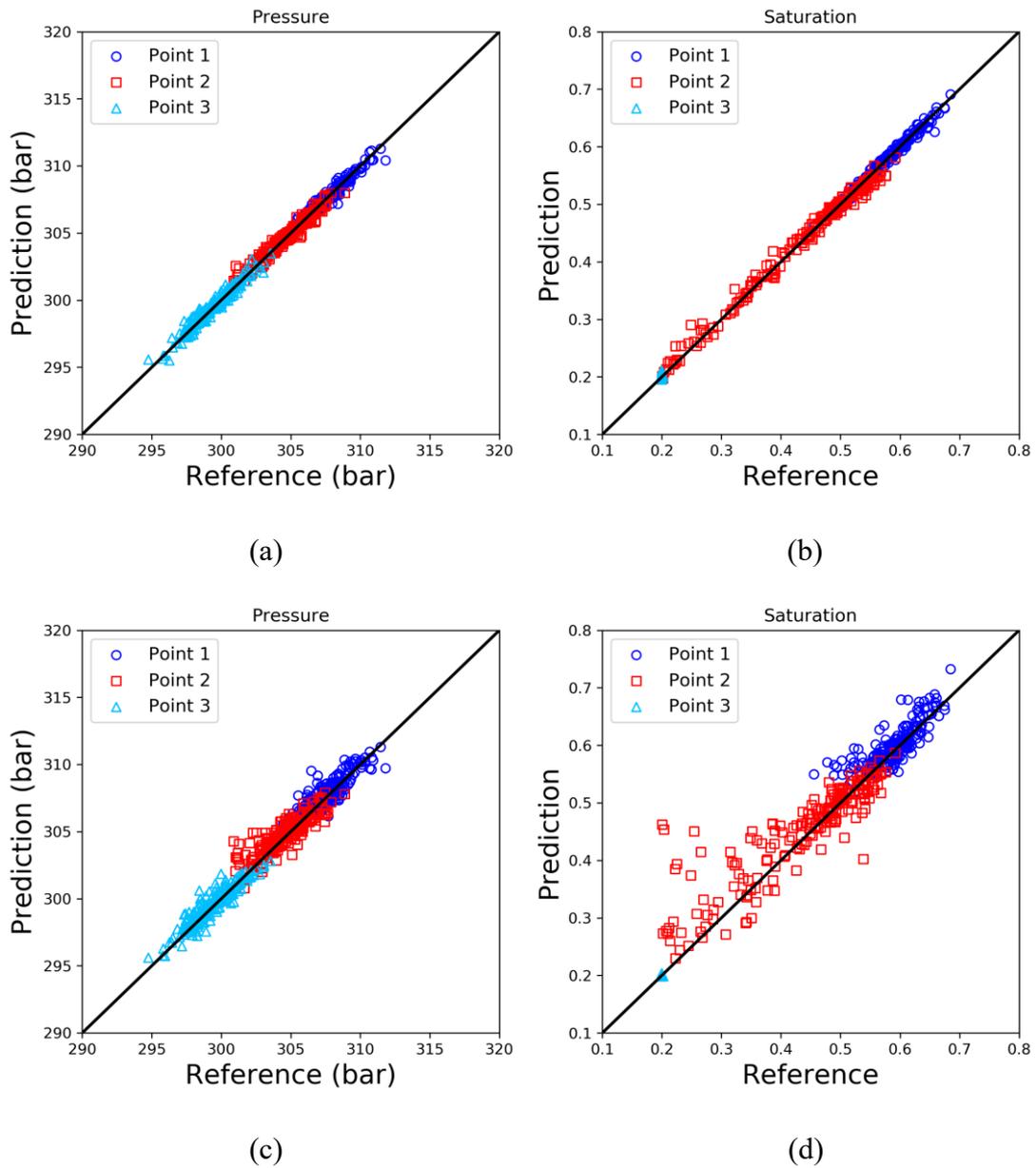

**Figure 5.** Scatter plots of reference values and predictions from TgCNN and CNN surrogates: (a) pressure with TgCNN; (b) saturation with TgCNN; (c) pressure with CNN; (d) saturation with CNN.

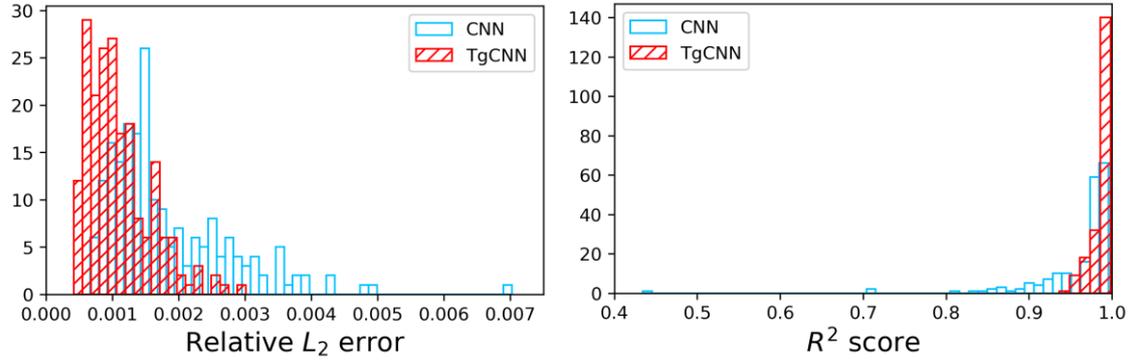

(a)

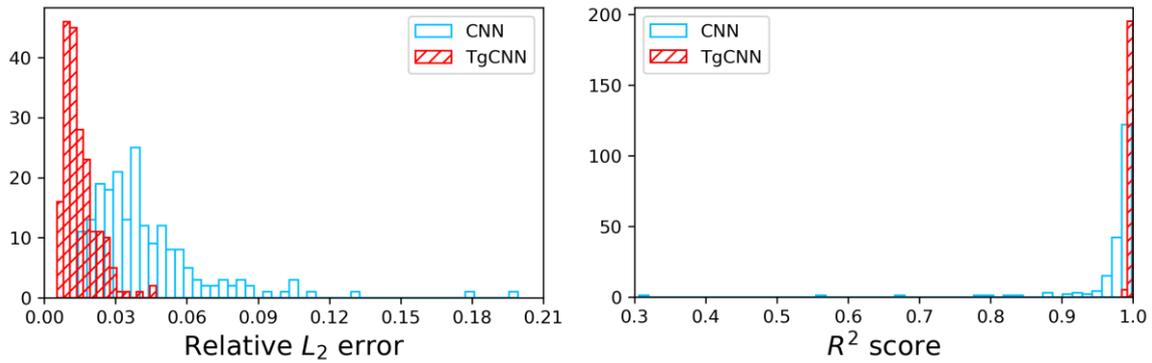

(b)

**Figure 6.** Histogram of relative $L_2$ error and $R^2$ score for (a) pressure and (b) saturation of 200 test realizations.

**4.3 Piecewise training strategy for varying well controls**

In this subsection, a new case with a more complicated simulation is studied, in which the well controls are altered during the production process and a piecewise training strategy for TgCNN is proposed. The simulation time of the previous case in section 4.2 is extended to 200 days in this case. At the end of the first 100 time-steps, the BHP of the producing well is switched to 250 bar and the BHP of the injection well is switched to 350 bar. Other settings are the same as the former case unless otherwise stated. Considering that the TgCNN surrogates have been trained for the first 100 time-steps, there is no need to retrain the TgCNN surrogates completely. New surrogates need to be constructed only for the extended time period, and stacked with the previously trained models. This constitutes an advantage of the piecewise training strategy, because the surrogates can be constructed in a dynamic manner with the development and management of oil fields. Moreover, the piecewise training strategy can also simplify the approximation for the nonlinear and discontinuous process with deep learning models, because it is easier to approximate each segment individually than to approximate the whole process. In this work, the total 200 time-steps are divided into two segments, with each having 100 time-steps. Since the surrogates of the first segment have already been trained in subsection 4.2, only the surrogates for the next segment need to be trained here. It is worth noting that the states at the last step of the first segment can serve as the initial conditions of the second segment.

In this case, we still use 80 realizations solved with the simulator as labeled training data, and 800 virtual realizations to impose physical constraints. It takes approximately 14.60 h (52559.26 s) to train the TgCNN surrogates for the second segment. Once trained, the surrogates of the two segments can be stacked together to predict solutions for the whole time-span of the models. The piecewise training strategy can also be adopted to construct the CNN surrogate models. New CNN surrogates are also trained for the second segment, and stacked together with CNN surrogates constructed in the former case. New test realizations can be generated and solved with the simulator to test the accuracy of the trained surrogates. The histograms of relative $L_2$ error and $R^2$ score of the predictions from TgCNN surrogates and CNN surrogates for the 200 test realizations are presented in **Figure 7**. Obviously, the TgCNN surrogates achieve higher accuracy than the CNN surrogates. The prediction results of a sampled realization are presented in **Figure 8**, and compared with the reference values from the simulator. It can be seen that sufficient prediction accuracy can be obtained with the piecewise trained TgCNN surrogates. Although there are some errors in the pressure predictions, they are still within the tolerance. Actually, the production data at well points are of greater concern, especially during the inverse modeling process, which can be calculated with the predicted pressure and saturation using Eq. (2). The predicted oil and water production rates of five production wells for the sampled realization are presented in **Figure 9** and **Figure 10**, and compared with the reference values from the reservoir simulator. Even though some details may not match perfectly, the overall predictions are acceptable, especially in terms of order of magnitude. Therefore, the trained TgCNN surrogates can be used for inverse modeling to improve efficiency by replacing the numerical solvers.

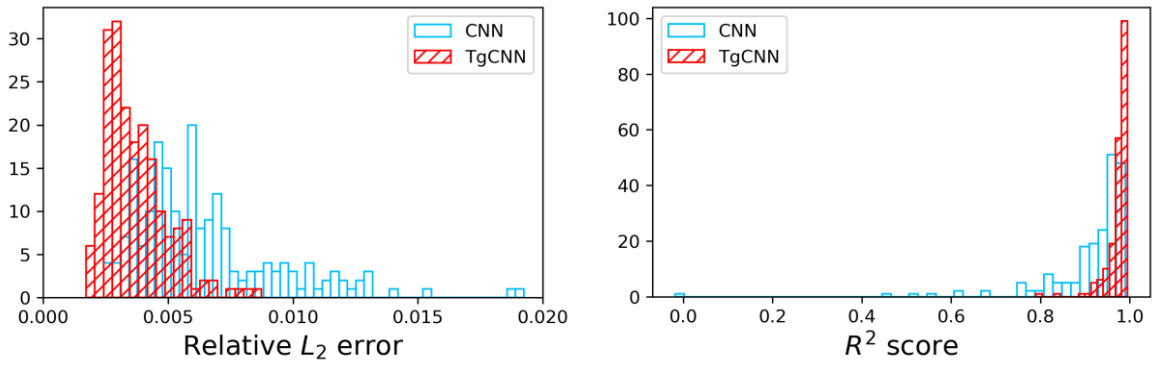

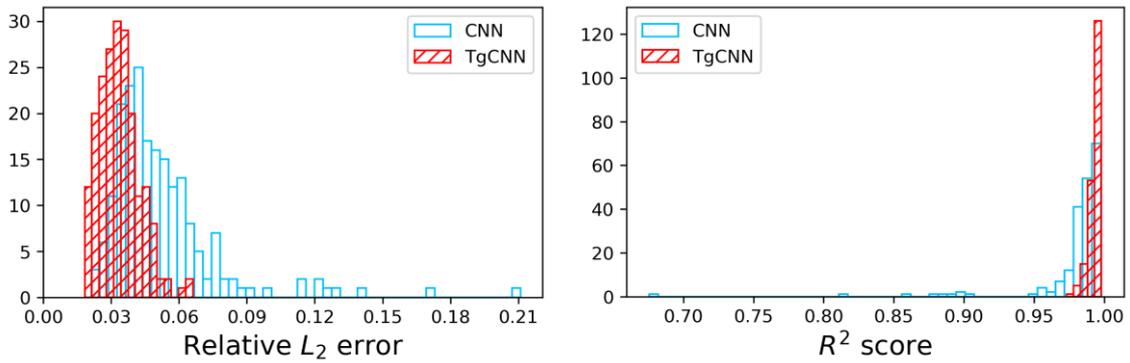

(b)

**Figure 7.** Histogram of relative $L_2$ error and $R^2$ score for (a) pressure and (b) saturation of 200 test realizations.

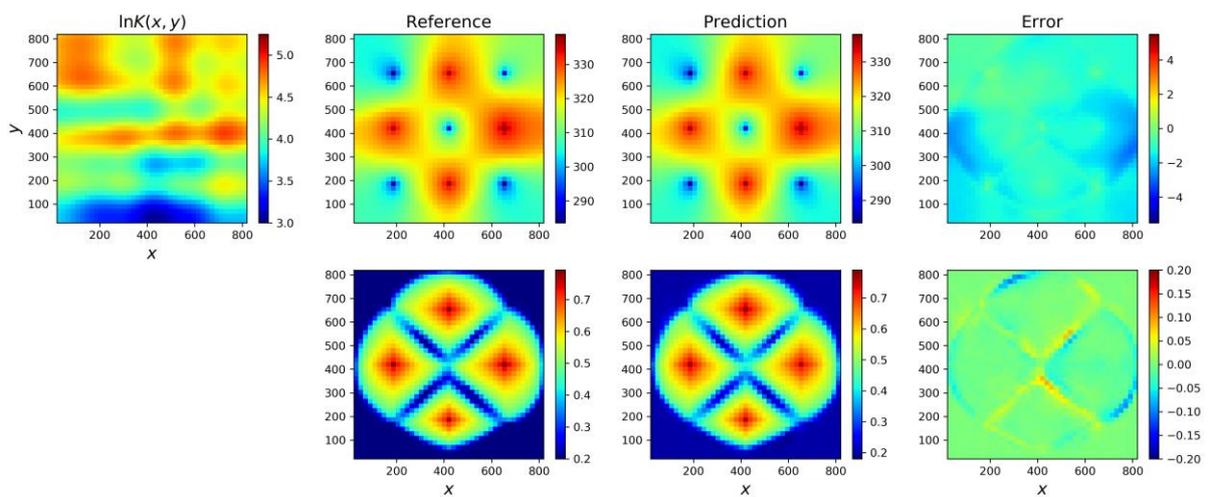

**Figure 8.** Predictions and references of pressure and saturation distribution for the sampled permeability field.

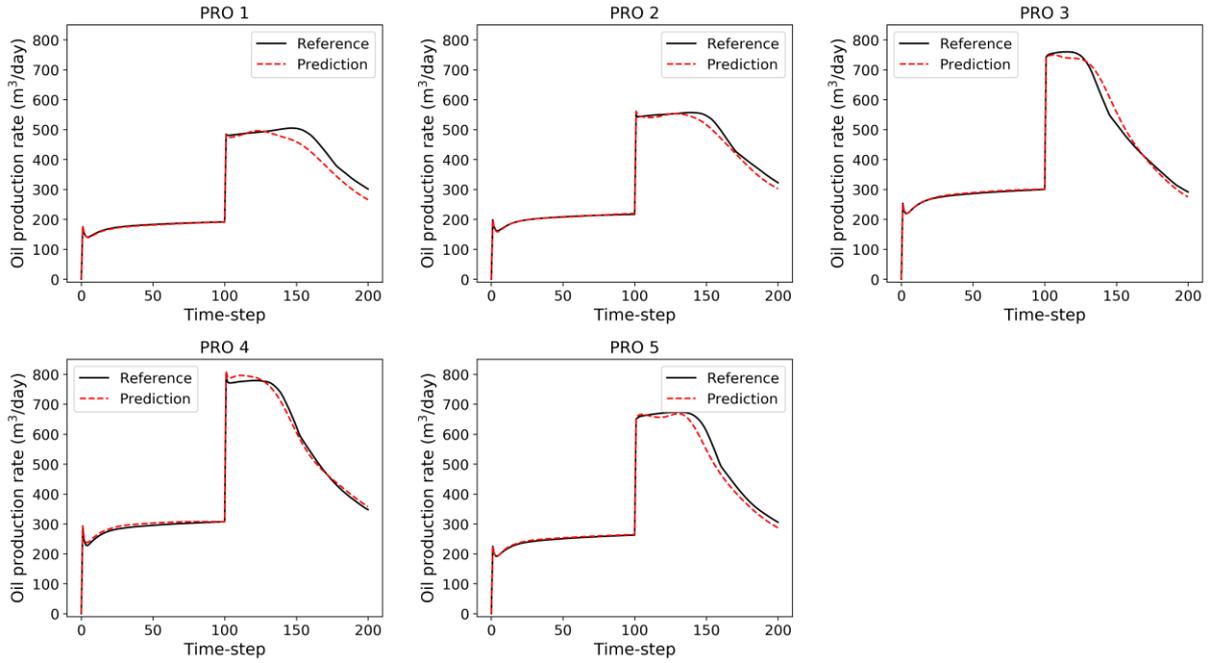

**Figure 9.** Predictions and reference values of oil production rates for the sampled realization.

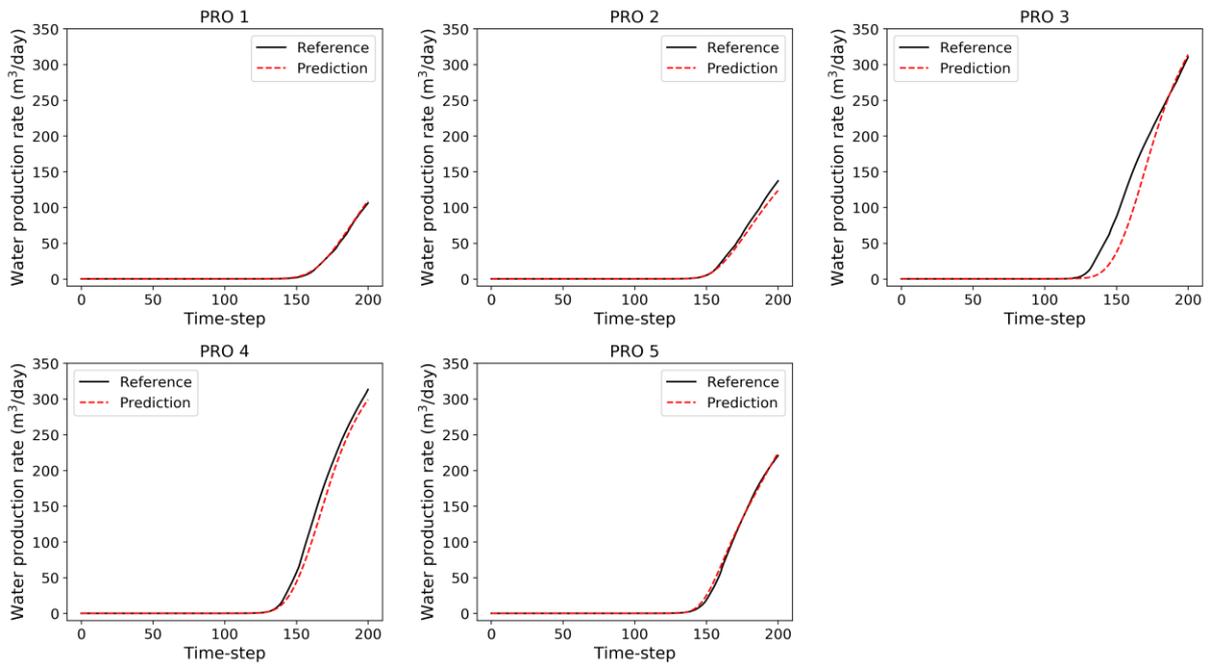

**Figure 10.** Predictions and reference values of water production rates for the sampled realization.

### 4.4 Increasing the variance of the stochastic permeability field

In this subsection, the variance of the stochastic permeability field is increased to investigate the performance of TgCNN for scenarios with stronger parameter heterogeneity. Still consider the case in subsection 4.1, but the standard deviation of the stochastic permeability field is increased to 0.8, i.e., $\sigma_{\ln K} = 0.8$. Due to the increased heterogeneity, more labeled data and virtual realizations are utilized to train the TgCNN surrogate here. In this case, 150 realizations are solved with the numerical simulator to provide labeled training data, and 1000 virtual realizations are generated to impose the theory constraints in the training process. It takes approximately 9.29 h (33437.77 s) to train the TgCNN surrogate. In addition, 200 test realizations are generated and solved with the numerical simulator to evaluate the accuracy of the trained surrogates. The correlation between the predictions from the surrogates and reference values from the simulator for three sample points of the 200 test realizations are presented in the scatter plots in **Figure 11**. The mean of relative $L_2$ error and $R^2$ score results for the 200 test realizations are listed in **Table 1**. The results demonstrate the effectiveness of the TgCNN for cases with larger parameter variance.

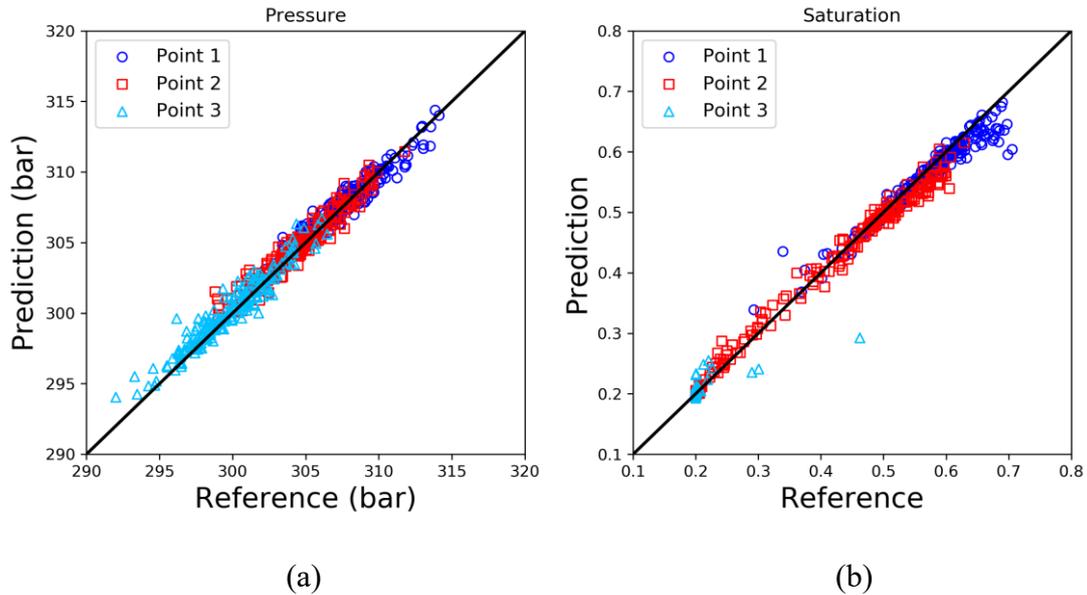

**Figure 11.** Scatter plots of reference values and predictions from TgCNN for (a) pressure and (b) saturation.

The transfer learning strategy can also be adopted for this case with the constructed models in subsection 4.2 being pretraining models, which has also been utilized in Wang et al. (2021c). The networks can be initialized with the parameters of TgCNN surrogates obtained for the case in subsection 4.2, and then be trained with the new datasets. With the already learned knowledge in the pretrained models, it would be much easier to converge to sufficient precision for the networks in the new scenarios. In this case, the transfer learning process for the new variance scenario only takes approximately 1.864 h (6712.113 s) for 100 epochs, which is much more efficient than retraining the models directly. The mean of relative $L_2$ error and $R^2$ score results for the 200 test realizations with the transfer learning strategy are listed in **Table 1**. It can be seen that using the transfer learning strategy can assist to achieve slightly higher accuracy than retraining the models directly with much higher efficiency.

**Table 1.** Mean of relative $L_2$ error and $R^2$ score results for the 200 test realizations with different strategies.

|  | Pressure | | Saturation | | Training time (s) |
| --- | --- | --- | --- | --- | --- |
|  | Relative $L_2$ error | $R^2$ score | Relative $L_2$ error | $R^2$ score |  |
| Retrain directly | 2.3669e-03 | 9.4692e-01 | 2.9224e-02 | 9.9096e-01 | 33437.765 |
| Transfer learning | 2.1744e-03 | 9.5354e-01 | 2.4570e-02 | 9.9368e-01 | 6712.113 |

## 5 Inverse Modeling with TgCNN-based IES

The constructed TgCNN surrogates are utilized for inverse modeling of the water flooding two-phase flow problem by combining them with IES in this section.

### 5.1 Inverse modeling with measurements of different time segments

In this work, the permeability fields are the model parameters to be estimated. The reference permeability field generated with KLE is presented in **Figure 12**, which is assumed to be unknown and needs to be estimated from the production measurements. The collected measurements include the oil and water production rate at the producing wells, as well as the water injection rate at the injecting wells. In IES, there are 100 realizations of permeability field in the ensemble, which would be updated simultaneously in the inversion process. The maximum number of iterations is set to be 10 while updating the parameters.

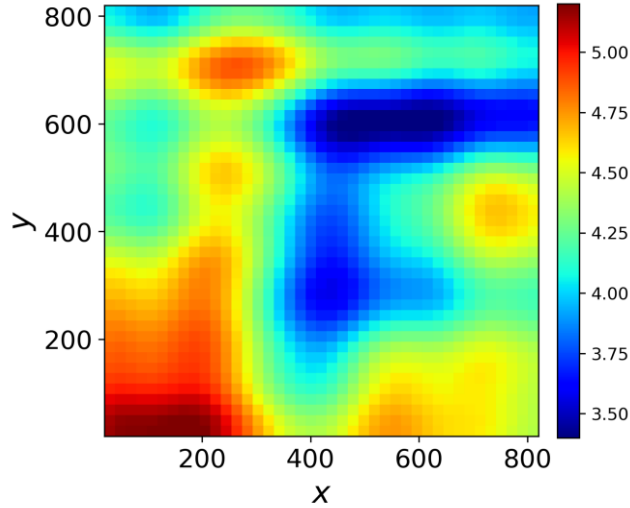

**Figure 12.** Reference permeability field.

First, the production data of the first 100 days are used for the inversion process. The mean of initial realizations and mean of estimated permeability fields are presented in **Figure 13** (a) and (b), which show that the estimated permeability is similar to the reference field with the root mean square error (RMSE) being 0.408 (**Table 2**). In addition, the uncertainty of the stochastic permeability field is reduced after the inversion process, especially at the well locations, as shown in **Figure 13** (d) and (e). The estimated permeability fields can be inputted into the simulator to predict the future states of the reservoir, and the total simulation time is extended to 400 days to test the prediction accuracy after inversion here. The production and injection data matching and future state prediction results are presented in **Figure 14**, **Figure 15**, and **Figure 16**. Obviously, the production curves of different realizations in the ensemble are diverse prior to the inversion process, and after assimilating the measurement data, the curves of different realizations converge to the reference values. The data matching and prediction results demonstrate the effectiveness of the TgCNN surrogate-based IES. The water front has not reached the producers for the first 100 days, and thus the water production rate is zero during the observation period, as shown in **Figure 15.** Even though the water production data are not utilized in this inversion process, relatively accurate predictions of future water production states can still be achieved.

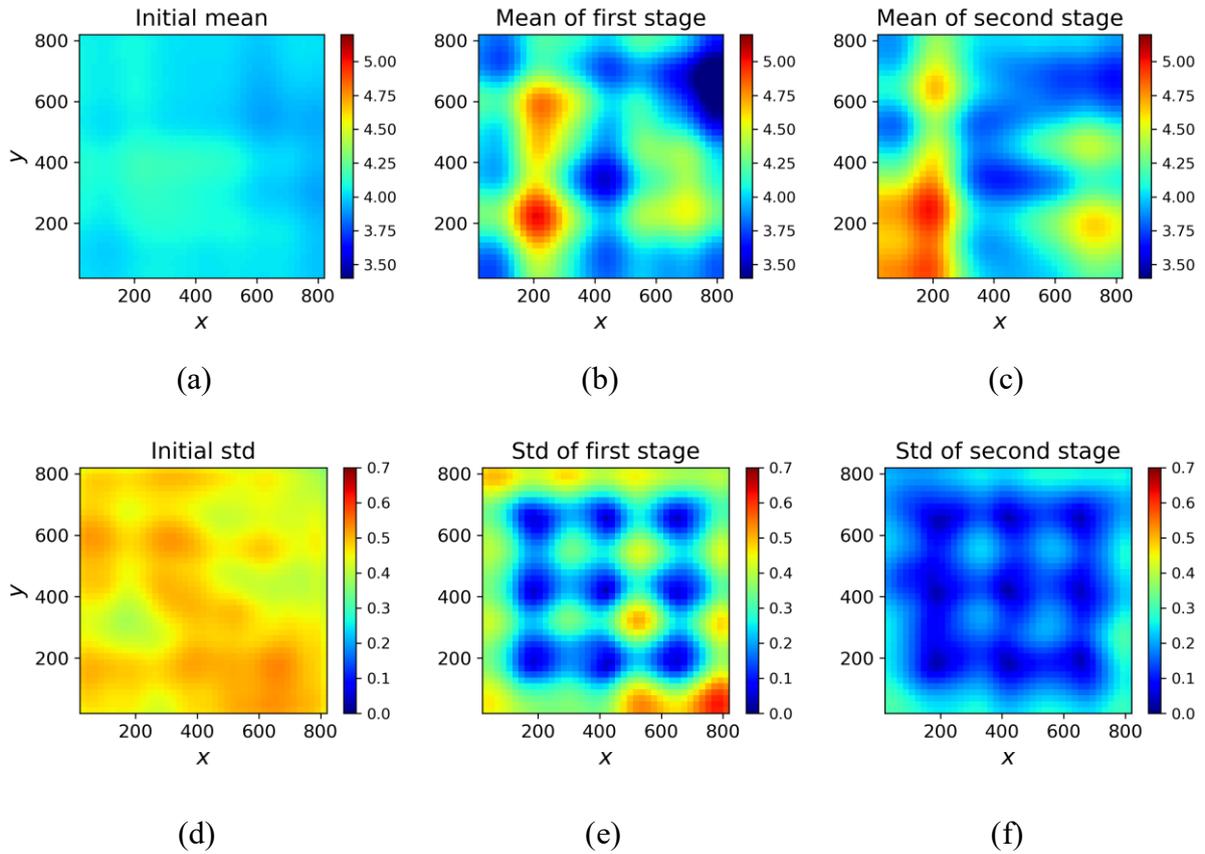

**Figure 13.** (a) Mean of initial realizations; (b) mean of estimated permeability fields with production data of the first 100 days; (c) mean of estimated permeability fields with production data of 200 days; (d) standard deviation of initial realizations; (e) standard deviation of estimated permeability fields with production data of the first 100 days; (f) standard deviation of estimated permeability fields with production data of 200 days.

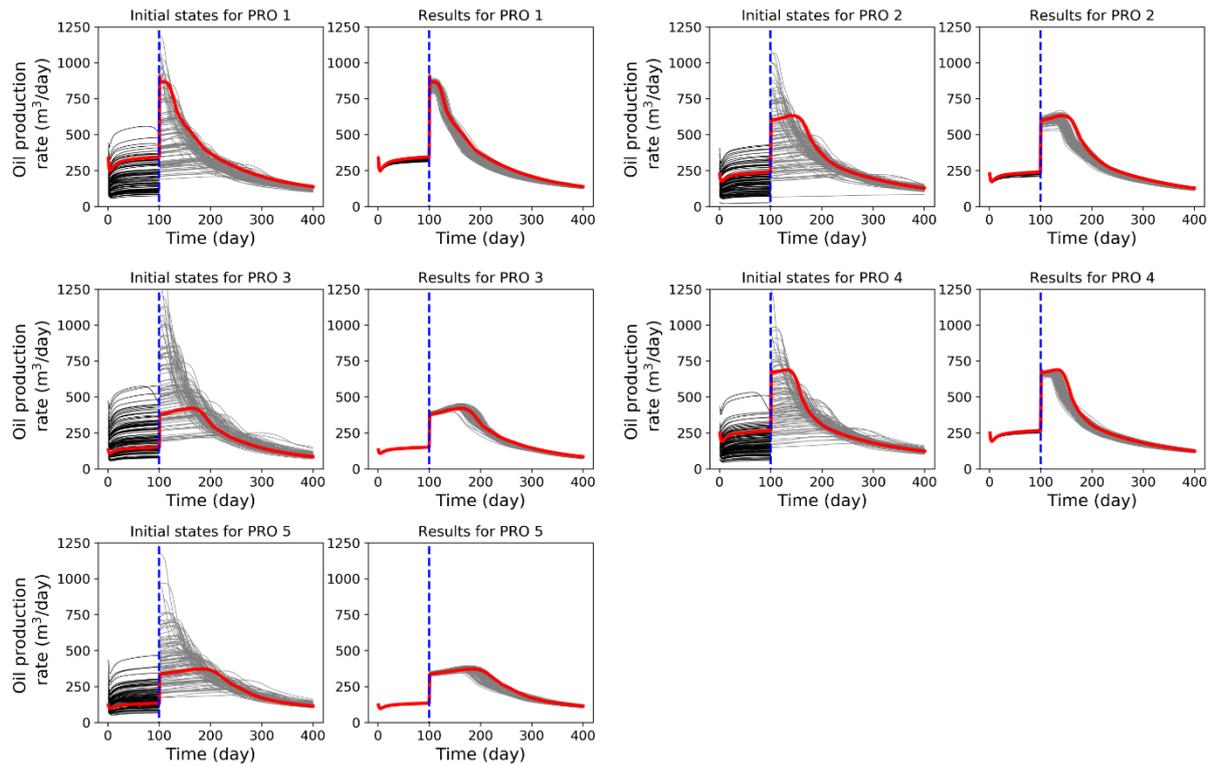

**Figure 14.** Oil production data matching and prediction results. The red lines denote the oil production rate for the reference realization. The blue dashed lines separate the data matching phase and the prediction phase. The black lines before the blue dashed lines denote the production rate of different realizations in the data matching phase, and the gray lines after the blue dashed lines denote the production rate of different realizations in the prediction phase.

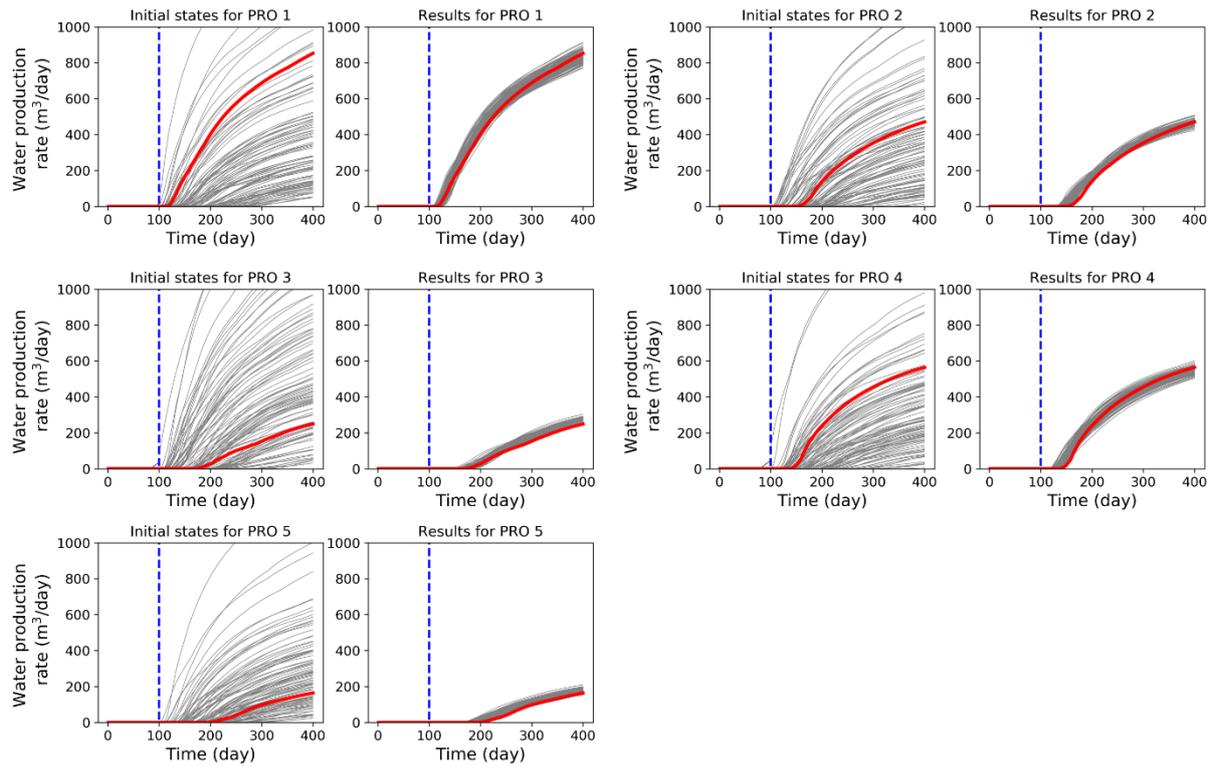

**Figure 15.** Water production data matching and prediction results. The red lines denote the water production rate for the reference realization. The blue dashed lines separate the data matching phase and the prediction phase. The black lines before the blue dashed lines denote the production rate of different realizations in the data matching phase, and the gray lines after the blue dashed lines denote the production rate of different realizations in the prediction phase.

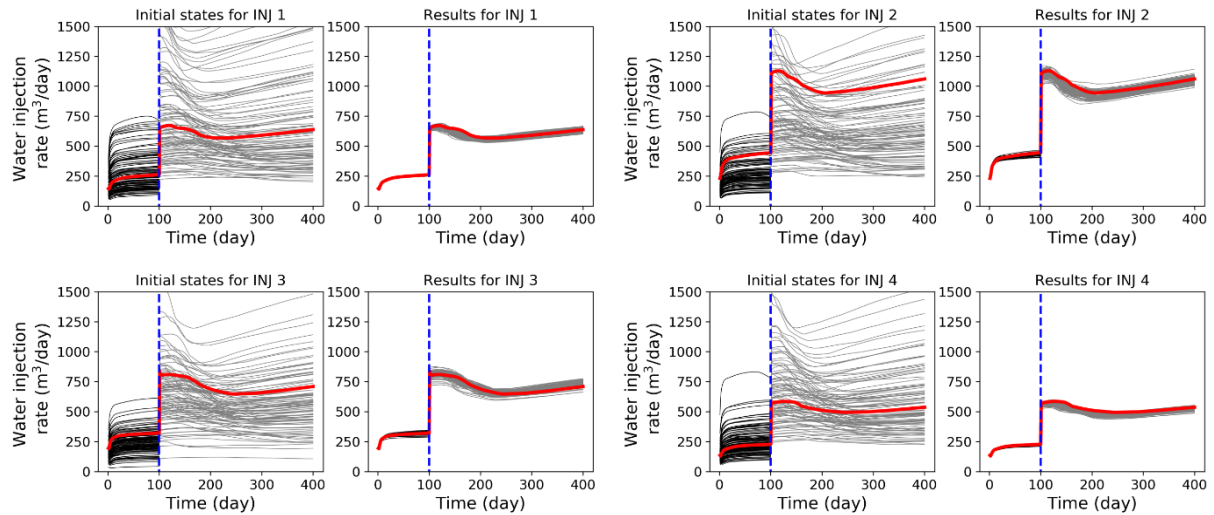

**Figure 16.** Water injection data matching and prediction results. The red lines denote the water injection rate for the reference realization. The blue dashed lines separate the data matching phase and the prediction phase. The black lines before the blue dashed lines denote the water injection rate of different realizations in the data matching phase, and the gray lines after the blue dashed lines denote the water injection rate of different realizations in the prediction phase.

With the development of oil reservoirs, the production data of the next 100 days can also be collected, which can then be used for further revision of previous inversion results. The trained piecewise TgCNN surrogates are utilized, and the estimation results of permeability are presented in **Figure 13** (c), which are much closer to the reference permeability field with the RMSE being 0.255 (**Table 2**). It is obvious that higher accuracy of estimation can be achieved with more available measurements. Of course, the inversion time has also increased due to the increased amount of measurement data, as shown in **Table 2**. Furthermore, compared to **Figure 13** (e), the standard deviation of the realizations has also been significantly reduced, as shown in **Figure 13** (f). The standard deviation of not only the area around the well points, but also the areas between wells, has decreased significantly, which means that the uncertainty of the unknown stochastic permeability field can be further reduced with more available measurements. Even though the production data come from the same well locations (i.e., the same spatial information), larger areas of the formation are affected in time. Therefore, the increase of measurement data in the time dimension can lead to better inversion of spatial model parameters. The data matching and prediction results are also presented in **Figure 17**, **Figure 18**, and **Figure 19**. It can be seen that the predictions are more accurate, and the uncertainty of the predictions is also decreased, when more collected measurements are assimilated for inverse modeling.

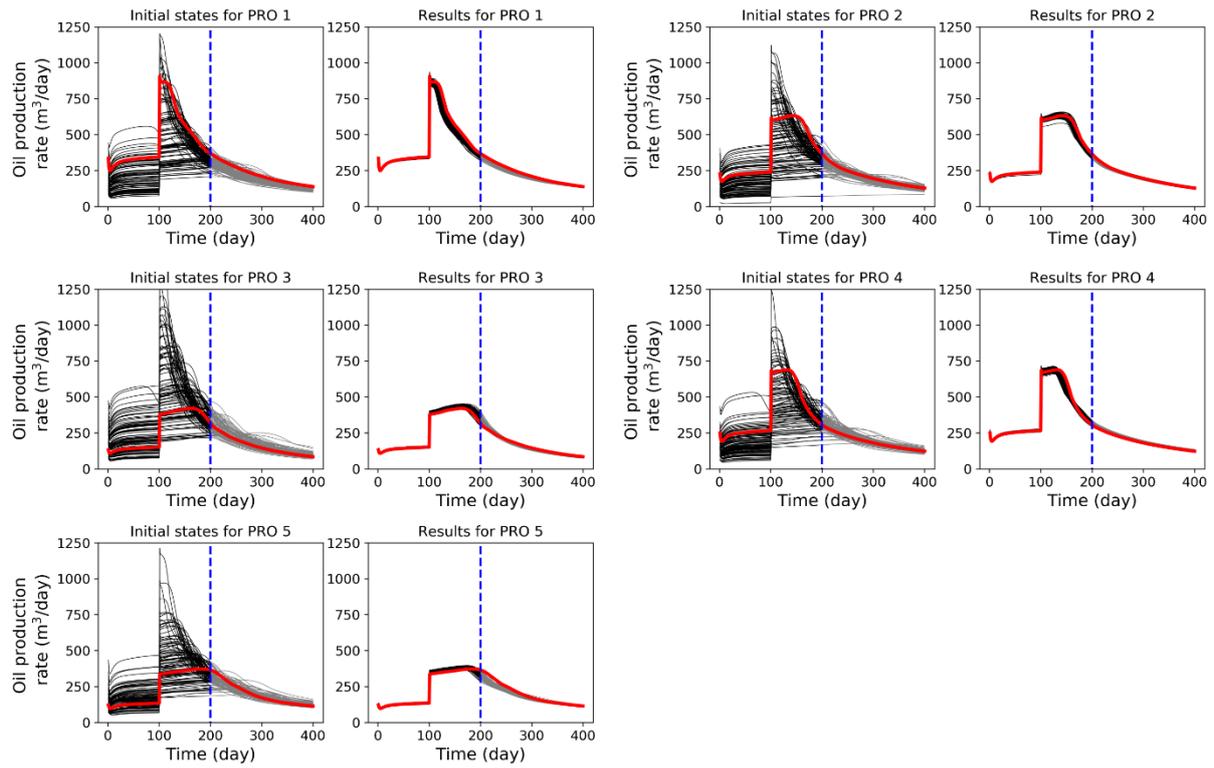

**Figure 17.** Oil production data matching and prediction results. The red lines denote the oil production rate for the reference realization. The blue dashed lines separate the data matching phase and the prediction phase. The black lines before the blue dashed lines denote the production rate of different realizations in the data matching phase, and the gray lines after the blue dashed lines denote the production rate of different realizations in the prediction phase.

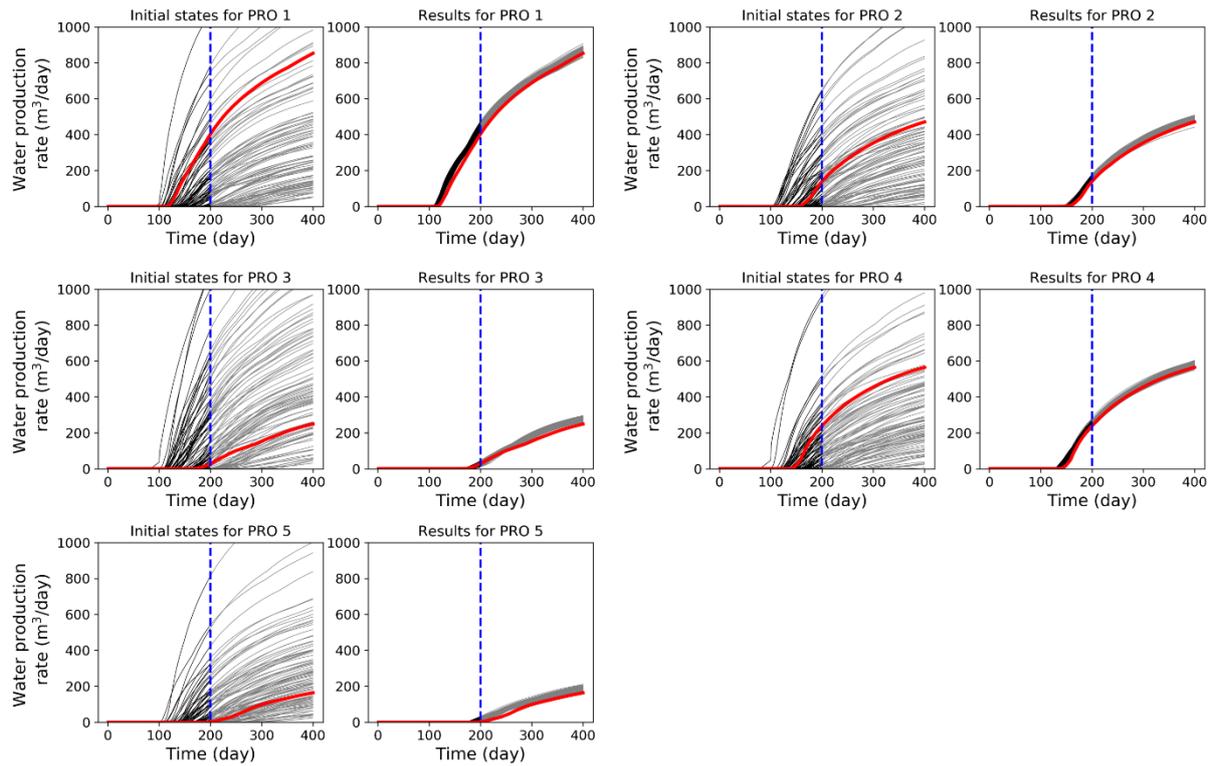

**Figure 18.** Water production data matching and prediction results. The red lines denote the water production rate for the reference realization. The blue dashed lines separate the data matching phase and the prediction phase. The black lines before the blue dashed lines denote the production rate of different realizations in the data matching phase, and the gray lines after the blue dashed lines denote the production rate of different realizations in the prediction phase.

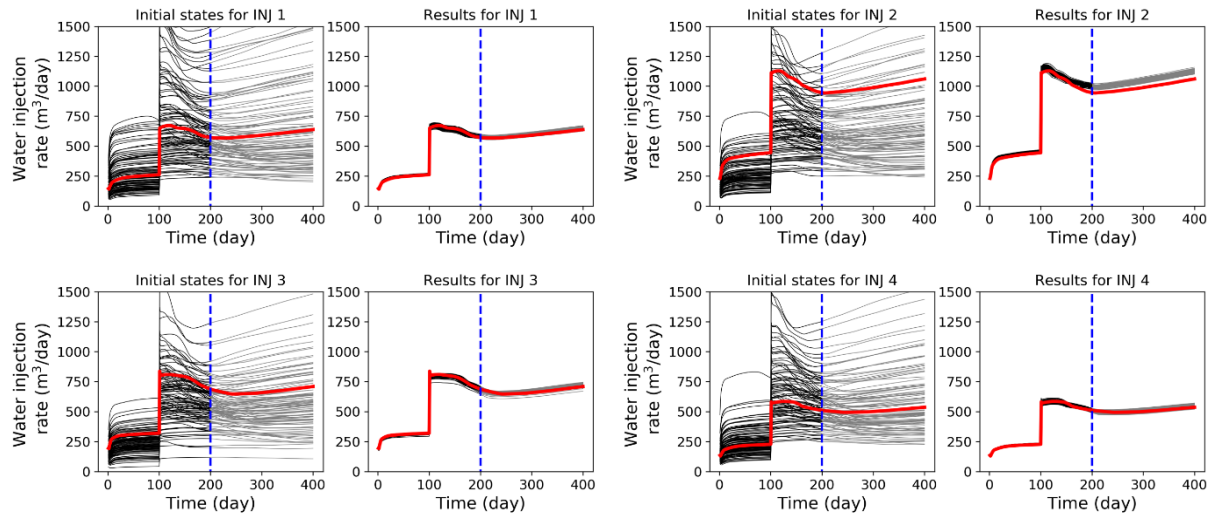

**Figure 19.** Water injection data matching and prediction results. The red lines denote the water injection rate for the reference realization. The blue dashed lines separate the data matching phase and the prediction phase. The black lines before the blue dashed lines denote the water injection rate of different realizations in the data matching phase, and the gray lines after the blue dashed lines denote the water injection rate of different realizations in the prediction phase.

Table 2. Inversion results with different amounts of measurement data.

| Data amount | Ensemble size | Iteration number | Inversion time (s) | RMSE |
|---|---|---|---|---|
| 100 days | 100 | 3 | 97.97 | 0.408 |
| 200 days | 100 | 3 | 352.12 | 0.255 |

### 5.2 Inverse modeling with unknown field variance

In this subsection, inverse modeling is performed for different reference cases of different variances. The variances of the underlying reference fields are unknown in the inversion process, which may, therefore, be different from those of the generated initial realizations in the TgCNN-based IES. Consider reference permeability field patterns similar to those in subsection 5.1, but generated with different standard deviations, as shown in **Figure 20** (a), in which the same scale of color bar is set to show the different variability of the fields. However, for all of these cases, the initial realizations in the TgCNN-based IES are the same as those used in subsection 5.1, which are generated with the standard deviation being 0.5. The measurements of the first 200 days are utilized for inversion. All of the other settings are the same as the former case, unless otherwise specified.

The inversion results for cases with different standard deviations are presented in **Figure 20** (b). It can be seen that the estimations of the permeability fields are similar to the respective reference fields with different standard deviations. In other words, even though the TgCNN surrogates are trained with realizations with the standard deviation being 0.5, they can still be used for inversion of cases with different standard deviations (0.2, 0.4, 0.6, 0.8, etc.), and achieve satisfactory inversion accuracy. The performance of the TgCNN surrogate-based IES seems to not be highly sensitive to the variance of the permeability fields, which shows the robustness of this inversion method. Of course, this insensitivity to variance is not infinite. Specifically, the performance of inversion would diminish as the real variance of the reference field gets further away from the variance of the realizations used to train the surrogates.

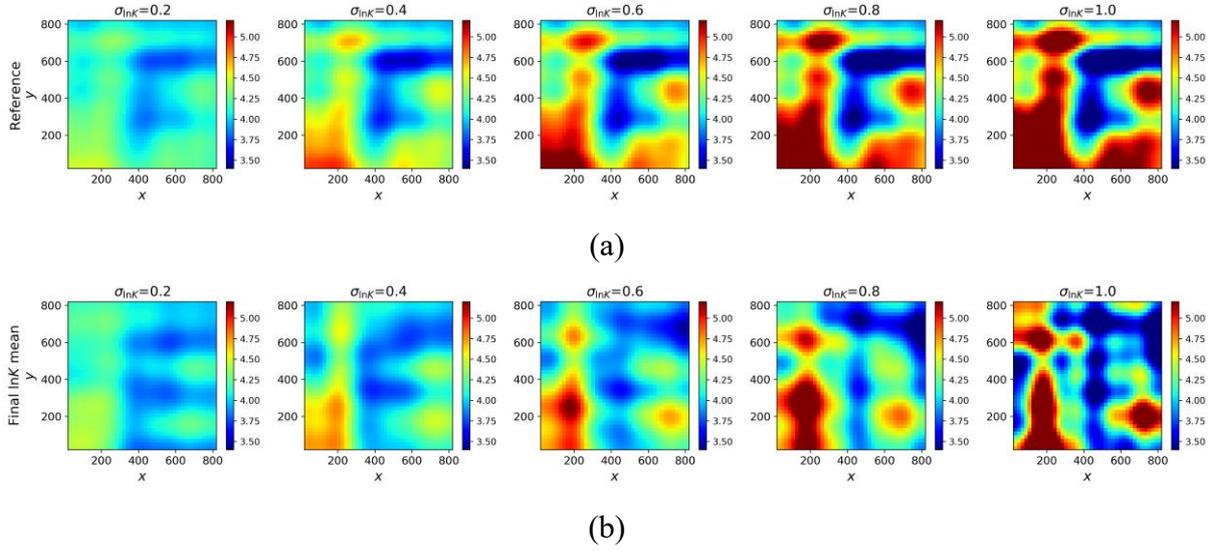

Figure 20. Reference fields (a) and inversion results (b) for cases with different standard deviations.

## 6 Discussions and Conclusions

In this work, the proposed TgCNN framework in Wang et al. (2021b) is extended to two-phase flow problems in porous media, and utilized for efficient inverse modeling. In two-phase flow problems, more complex governing equations than that of single-phase flow need to be considered (oil phase and water phase), and more state variables need to be approximated (pressure and saturation), which means that more than one network need to be constructed. In order to achieve theory-guided training, the governing equations are discretized with the FD method and transformed into residual forms. The residuals are then incorporated into the loss function of the TgCNN models. Moreover, the networks of different variables are coupled with each other by the governing equations, and need to be trained simultaneously. The trained TgCNN surrogates can be further used for efficient inverse modeling in two-phase flow problems by replacing the numerical flow simulators for faster forward calculation.

A two-dimensional water flooding problem (oil-water flow) is considered to test the performance of the TgCNN framework in two-phase flow problems. With the same amount of labeled training data, TgCNN surrogates achieve much better accuracy than ordinary CNN surrogates, which demonstrates the effectiveness of theory-guidance in the training process. Furthermore, for the nonlinear and discontinuous well control varying situation in the production process, the piecewise training strategy is proposed, in which the whole time-span is divided into several segments, and the TgCNN surrogates are constructed for the different segments, respectively. The results show that the piecewise trained TgCNN surrogates can achieve sufficient accuracy for those discontinuous well control processes. For scenarios with larger parameter variance, the TgCNN surrogates can also obtain satisfactory performance. In addition, the transfer learning strategy can be adopted to leverage the learned information of pretrained models and accelerate the convergence process. The trained TgCNN surrogates are further used for inverse modeling with the IES algorithm, and satisfactory estimation results for the permeability field are achieved. Accurate data matching and prediction results of production states can also be obtained efficiently with the TgCNN surrogate-based IES. The results also demonstrate that sensitivity towards the variance of the stochastic field of the inversion method is not very high, which makes it usable for scenarios with different field variances.

**Acknowledgements**

This work is partially funded by the Shenzhen Key Laboratory of Natural Gas Hydrates (Grant No. ZDSYS20200421111201738) and the SUSTech - Qingdao New Energy Technology Research Institute. All of the data of the figures shown in this study are made available for download from a public repository of research data through the following URL: https://figshare.com/articles/dataset/TgCNN_for_two-phase_flow/16752064 (https://doi.org/10.6084/m9.figshare.16752064.v1).

**References**

Bottou, L. (2010). *Large-Scale Machine Learning with Stochastic Gradient Descent*. Paper presented at the Proceedings of COMPSTAT'2010, Heidelberg. Physica-Verlag HD.
Chang, H., Liao, Q., & Zhang, D. (2017). Surrogate model based iterative ensemble smoother for subsurface flow data assimilation. *Advances in Water Resources, 100*, 96-108. http://www.sciencedirect.com/science/article/pii/S0309170816307345

Chen, Y., & Oliver, D. S. (2013). Levenberg–Marquardt forms of the iterative ensemble smoother for efficient history matching and uncertainty quantification. *Computational Geosciences, 17*(4), 689-703. https://doi.org/10.1007/s10596-013-9351-5

Hamdi, H., Hajizadeh, Y., & Sousa, M. C. (2015). *Gaussian Process for Uncertainty Quantification of Reservoir Models*. Paper presented at the SPE/IATMI Asia Pacific Oil & Gas Conference and Exhibition. https://doi.org/10.2118/176074-MS

Karumuri, S., Tripathy, R., Bilionis, I., & Panchal, J. (2020). Simulator-free solution of high-dimensional stochastic elliptic partial differential equations using deep neural networks. *Journal of Computational Physics, 404*, 109120. https://www.sciencedirect.com/science/article/pii/S0021999119308253

Kingma, D. P., & Ba, J. L. (2015). *Adam: A Method for Stochastic Optimization*. Paper presented at the International Conference on Learning Representations.

Li, G., & Reynolds, A. C. (2009). Iterative Ensemble Kalman Filters for Data Assimilation. *SPE Journal, 14*(03), 496-505. https://doi.org/10.2118/109808-PA

Li, H., Chang, H., & Zhang, D. (2009). *Stochastic Collocation Methods for Efficient and Accurate Quantification of Uncertainty in Multiphase Reservoir Simulations*. Paper presented at the SPE Reservoir Simulation Symposium. https://doi.org/10.2118/118964-MS

Li, H., & Zhang, D. (2007). Probabilistic collocation method for flow in porous media: Comparisons with other stochastic methods. *Water Resources Research, 43*(9). https://doi.org/10.1029/2006WR005673

Liao, Q., Zeng, L., Chang, H., & Zhang, D. (2019). Efficient History Matching Using the Markov-Chain Monte Carlo Method by Means of the Transformed Adaptive Stochastic Collocation Method. *SPE Journal, 24*(04), 1468-1489. https://doi.org/10.2118/194488-PA

Mo, S., Zhu, Y., Zabaras, N., Shi, X., & Wu, J. (2019). Deep Convolutional Encoder-Decoder Networks for Uncertainty Quantification of Dynamic Multiphase Flow in Heterogeneous Media. *Water Resources Research, 55*(1), 703-728. https://doi.org/10.1029/2018WR023528

Oliver, D. S., Reynolds, A. C., & Liu, N. (2008). *Inverse Theory for Petroleum Reservoir Characterization and History Matching*: Cambridge University Press.

Peaceman, D. W. (1983). Interpretation of Well-Block Pressures in Numerical Reservoir Simulation With Nonsquare Grid Blocks and Anisotropic Permeability. *SPE Journal, 23*(03), 531-543. https://doi.org/10.2118/10528-PA

Raissi, M., Perdikaris, P., & Karniadakis, G. E. (2019). Physics-informed neural networks: A deep learning framework for solving forward and inverse problems involving nonlinear partial differential equations. *Journal of Computational Physics, 378*, 686-707. https://www.sciencedirect.com/science/article/pii/S0021999118307125

Rana, S., Ertekin, T., & King, G. R. (2018). *An Efficient Probabilistic Assisted History Matching Tool Using Gaussian Processes Proxy Models: Application to Coalbed Methane Reservoir*. Paper presented at the SPE Annual Technical Conference and Exhibition. https://doi.org/10.2118/191655-MS

Schlumberger. (2009). *Eclipse Reference Manual 2009.1*. Houston.

Tang, M., Liu, Y., & Durlofsky, L. J. (2020). A deep-learning-based surrogate model for data assimilation in dynamic subsurface flow problems. *Journal of Computational Physics, 413*, 109456. https://www.sciencedirect.com/science/article/pii/S0021999120302308

Tripathy, R. K., & Bilionis, I. (2018). Deep UQ: Learning deep neural network surrogate models for high dimensional uncertainty quantification. *Journal of Computational Physics, 375*, 565-588. https://www.sciencedirect.com/science/article/pii/S0021999118305655

Wang, N., Chang, H., & Zhang, D. (2021a). Deep-Learning-Based Inverse Modeling Approaches: A Subsurface Flow Example. *Journal of Geophysical Research: Solid Earth, 126*(2), e2020JB020549. https://doi.org/10.1029/2020JB020549

Wang, N., Chang, H., & Zhang, D. (2021b). Efficient Uncertainty Quantification and Data Assimilation via Theory-Guided Convolutional Neural Network. *SPE Journal*, 1-29. https://doi.org/10.2118/203904-PA

Wang, N., Chang, H., & Zhang, D. (2021c). Efficient uncertainty quantification for dynamic subsurface flow with surrogate by Theory-guided Neural Network. *Computer Methods in Applied Mechanics and Engineering, 373*, 113492. http://www.sciencedirect.com/science/article/pii/S0045782520306770

Wang, N., Chang, H., & Zhang, D. (2021d). Theory-guided Auto-Encoder for surrogate construction and inverse modeling. *Computer Methods in Applied Mechanics and Engineering, 385*, 114037. https://www.sciencedirect.com/science/article/pii/S0045782521003686

Wang, N., Zhang, D., Chang, H., & Li, H. (2020). Deep learning of subsurface flow via theory-guided neural network. *Journal of Hydrology, 584*, 124700. http://www.sciencedirect.com/science/article/pii/S0022169420301608

Zeng, L., Chang, H., & Zhang, D. (2011). A Probabilistic Collocation-Based Kalman Filter for History Matching. *SPE Journal, 16*(02), 294-306. https://doi.org/10.2118/140737-PA

Zhang, D., Li, H., & Chang, H. (2011). *History Matching for Non-Gaussian Random Fields Using the Probabilistic Collocation Based Kalman Filter*. Paper presented at the SPE Reservoir Simulation Symposium. https://doi.org/10.2118/141893-MS

Zhong, Z., Sun, A. Y., Ren, B., & Wang, Y. (2021). A Deep-Learning-Based Approach for Reservoir Production Forecast under Uncertainty. *SPE Journal, 26*(03), 1314-1340. https://doi.org/10.2118/205000-PA

Zhu, Y., Zabaras, N., Koutsourelakis, P.-S., & Perdikaris, P. (2019). Physics-constrained deep learning for high-dimensional surrogate modeling and uncertainty quantification without labeled data. *Journal of Computational Physics, 394*, 56-81. https://www.sciencedirect.com/science/article/pii/S0021999119303559